\documentclass[10pt,journal,letterpaper]{IEEEtran}

\usepackage{epstopdf}
\usepackage{graphicx}
\usepackage{stmaryrd}
\usepackage{amsmath,amsthm,amssymb}
\usepackage{multirow,url}
\usepackage{color,cite}
\usepackage{algorithm}
\usepackage{algorithmicx}
\usepackage{algpseudocode}
\usepackage{setspace}

\newtheorem{thm}{Theorem}
\newtheorem{defn}{Definition}
\newtheorem{cor}{Corollary}

\newcommand{\eg}{\emph{e.g., }}

\ifodd 0
\newcommand{\rev}[1]{{\color{blue}#1}} 
\newcommand{\com}[1]{\textbf{\color{red} (COMMENT: #1)}} 
\newcommand{\comg}[1]{\textbf{\color{green} (COMMENT: #1)}}
\newcommand{\response}[1]{\textbf{\color{magenta} (RESPONSE: #1)}} 
\else
\newcommand{\rev}[1]{#1}
\newcommand{\com}[1]{}
\newcommand{\comg}[1]{}
\newcommand{\response}[1]{}
\fi

\begin{document}
\title{Evolutionarily Stable Spectrum Access}

\author{Xu Chen, \emph{Student Member, IEEE,} and Jianwei Huang, \emph{Senior Member, IEEE}\thanks{The authors are with the Network Communications and Economics Lab, Department of Information Engineering, The Chinese University of Hong Kong, Shatin, NT, Hong Kong. Emails: \{cx008,jwhuang\}@ie.cuhk.edu.hk.}}


%

\maketitle

\allowdisplaybreaks

\begin{abstract}
In this paper, we design distributed spectrum access mechanisms with both complete and incomplete network information. We propose an evolutionary spectrum access mechanism  with complete network information, and show that the mechanism achieves an equilibrium that is globally evolutionarily stable. With incomplete network information,  we propose a distributed learning mechanism, where each user utilizes local observations to estimate the expected throughput and learns to adjust its spectrum access strategy adaptively over time. We show that the learning mechanism converges to the same evolutionary equilibrium on the time average. Numerical results show that the proposed mechanisms achieve up to $35\%$ performance improvement over the distributed reinforcement learning mechanism in the literature, and are
robust to the perturbations of users' channel selections.
\end{abstract} 
\section{Introduction}
Cognitive radio is envisioned as a promising technique
to alleviate the problem of spectrum under-utilization \cite{ikey-1}.
It enables unlicensed wireless users (secondary users) to
opportunistically access the licensed channels owned by spectrum holders (primary users), and thus can significantly
improve the spectrum efficiency \cite{ikey-2}.

A key challenge of the cognitive radio technology is how
to share the spectrum resources efficiently in a distributed fashion. A common modeling approach is to consider selfish secondary
users, and model their interactions as \emph{non-cooperative games} (\eg \cite{Chen2012Spatial,ikey-20,ikey-15,ikey-22,ikey-21,ikey-16}).
%
%
Liu and Wu in \cite{ikey-15} modeled the interactions among spatially separated secondary users as congestion games with resource reuse. Elias \emph{et al.} in \cite{ikey-22} studied the competitive spectrum access by multiple interference-sensitive secondary users. Nie and Comniciu in \cite{ikey-21} designed a self-enforcing distributed spectrum access mechanism based on potential games.
Law \emph{et al.} in \cite{ikey-16} studied the price of anarchy of spectrum access game, and showed that users' selfish choices may significantly degrade system performance. A common assumption of the above results is that each user knows the complete network information. This is, however, often expensive or infeasible to achieve due to significant signaling overhead and the  competitiors' unwillingness to share information.


Another common assumption of all the above work is that secondary users are \emph{fully rational} and thus often adopt their channel selections based on best responses, i.e., the best choices they can compute by having the complete network information. To have full rationality, a user needs to have a high computational power to collect and analyze the network information in order to predict other users' behaviors. This is often not feasible due to the limitations of today's wireless devices.

Another body of related work focused on the design of spectrum access mechanisms assuming \emph{bounded rationality} of secondary users, i.e., each user tries to improve its strategy adaptively over time. Chen and Huang in \cite{Chen2011} designed an imitation-based spectrum access mechanism by letting secondary users imitate other users' successful channel selections. When not knowing the channel selections of other users, secondary users need to learn the environment and adapt the channel selection decisions accordingly. Authors in \cite{ikey-25,ikey-26} used no-regret learning to solve this problem, assuming that the users' channel selections are common information. The learning converges to a correlated equilibrium \cite{ikey-18}, wherein the common observed history serves as a signal to coordinate all users' channel selections.
%
%
When users' channel selections are not observable, authors in \cite{ikey-27,ikey-28,ikey-29} designed multi-agent multi-armed bandit learning algorithm to minimize the expected performance loss of distributed spectrum access. Li in \cite{ikey-9} applied reinforcement learning to analyze Aloha-type spectrum access.

In this paper, we propose a new framework of distributed spectrum access with and without complete network information (i.e., channel statistics and user selections). The common characteristics of algorithms under this framework is also \emph{bounded rationality}, which requires much less computation power than the full rationality case, and thus may better match the reality of wireless communications. We first propose an evolutionary game approach for distributed spectrum access with the complete network information, where each secondary user takes a \emph{comparison} strategy (i.e., comparing its payoff with the system average payoff) to evolve its spectrum access decision over time. We then propose a  learning mechanism for
distributed spectrum access with incomplete information, which does not require any prior knowledge of channel statistics or information exchange among users. In this case,
each secondary user estimates its expected throughput locally, and \emph{learns} to adjust its channel selection strategy adaptively.

The main results and contributions of this paper are as follows:
\begin{itemize}
\item \emph{Evolutionary spectrum access mechanism}: we formulate the distributed
spectrum access over multiple heterogeneous time-varying licensed channels as an evolutionary
spectrum access game, and study the evolutionary dynamics of spectrum access.
\item \emph{Evolutionary dynamics and stability}:
we show that the evolutionary spectrum access mechanism converges to the evolutionary equilibrium, and prove that it is globally evolutionarily stable.
\item \emph{Learning mechanism with incomplete information}: we further propose a learning mechanism
without the knowledge of channel statistics and user information exchange. We show that the learning mechanism converges to the
same evolutionary equilibrium on the time average.
\item \emph{Superior performance}: we show that the proposed mechanisms can achieve up to $35\%$ performance improvement over the distributed reinforcement learning mechanism in literature, and are
robust to the perturbations of users' channel selections.
\end{itemize}

%

The rest of the paper is organized as follows. We introduce the system
model in Section \ref{sec:System-Model}. After briefly reviewing the evolutionary game theory in Section \ref{sec:Overview-of-Evolutionary}, we present the evolutionary spectrum access mechanism with complete information in Section \ref{sec:Evolutionary-Game-Approach}. Then we introduce the learning mechanism in Section \ref{sec:Evolutionarily-Stable-Learning}. We illustrate
the performance of the proposed mechanisms through numerical results
in Section \ref{sec:Simulation-Results} and finally conclude in Section \ref{sec:Conclusion}.

\section{\label{sec:System-Model}System Model}
\begin{figure}[t]
\centering
\includegraphics[scale=0.6]{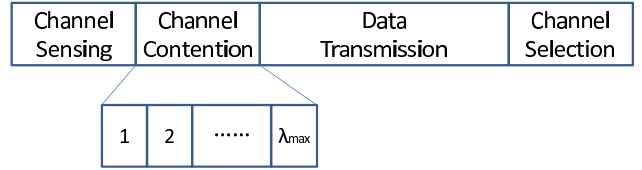}
\caption{\label{fig:Time-slot-structure} Multiple stages in a single time slot.}
\end{figure}

We consider a cognitive radio network with a set $\mathcal{M}=\{1,2,...,M\}$
of independent and\emph{ stochastically heterogeneous} licensed channels.
A set $\mathcal{N}=\{1,2,...,N\}$ of secondary users
try to opportunistically access these channels, when the channels
are not occupied by primary (licensed) transmissions. The system model
has a slotted transmission structure as in Figure \ref{fig:Time-slot-structure} and is described as follows.
\begin{itemize}
\item \emph{Channel State}: the channel state for a channel $m$ at time slot $t$ is \[
S_{m}(t)=\begin{cases}
0, & \mbox{if channel $m$ is occupied by}\\
& \mbox{primary transmissions,}\\
1, & \mbox{if channel $m$ is idle.}
\end{cases}\]


\item \emph{Channel State Changing}: for a channel $m$, we assume that
the channel state is an i.i.d. Bernoulli random variable, with an
idle probability $\theta_{m}\in(0,1)$ and a busy probability $1-\theta_{m}$.
This model can be a good approximation of the reality if the time slots
for secondary transmissions are sufficiently long or the primary transmissions
are highly bursty \cite{key-27}. Numerical results show that the proposed mechanisms also work well in the Markovian channel environment.

\item \emph{Heterogeneous Channel Throughput}: if a channel $m$ is idle, the achievable data rate $b_{m}(t)$ by a secondary user
 in each time slot $t$ evolves according
to an i.i.d. random process with a mean $B_{m}$, due to the local
environmental effects such fading. For example, in a frequency-selective Rayleigh fading
channel environment we can compute the channel data rate according to the Shannon capacity with the channel gain at a time slot being a realization
of a random variable that follows the exponential distribution \cite{rappaport1996wireless}.
\item \emph{Time Slot Structure}: each secondary user $n$ executes the
following stages synchronously during each time slot:
\begin{itemize}
\item \emph{Channel Sensing}: sense one of the channels based on the channel
selection decision generated at the end of previous time slot. Access the channel if it is idle.
\item \emph{Channel Contention}: use a backoff mechanism to resolve collisions when
multiple secondary users access the same idle channel. The contention stage of a time slot is divided into $\lambda_{\max}$
mini-slots\footnote{\rev{For the ease of exposition, we assume that the contention backoff size $\lambda_{\max}$ is fixed. This corresponds to an equilibrium model for the case that the backoff size $\lambda_{\max}$ can be dynamically tuned according to the 802.11 distributed coordination
function \cite{bianchi2000performance}. Also, we can enhance the performance of the backoff mechanism by determining optimal fixed contention backoff size according to the method in \cite{kriminger2011markov}.}} (see Figure~\ref{fig:Time-slot-structure}), and
user $n$ executes the following two steps.
\emph{First}, count down according to a randomly and uniformly chosen integral backoff
time (number of mini-slots) $\lambda_{n}$ between $1$ and $\lambda_{\max}$. \emph{Second,} once the timer expires, transmit RTS/CTS messages if the channel is clear (i.e., no ongoing transmission).
Note that if multiple users choose the same backoff value $\lambda_{n}$, a collision
will occur with RTS/CTS transmissions and no users can successfully grab the channel.
\item \emph{Data Transmission}:  transmit data packets if the RTS/CTS message exchanges go through and the user successfully
grabs the channel.
\item \emph{Channel Selection}: in the complete information case, users broadcast the chosen channel IDs to other users through a common
control channel\footnote{\rev{Please refer to \cite{lo2011survey} for the details on how to set up and maintain
a reliable common control channel in cognitive radio networks.}}, and then make the channel selection decisions based on the evolutionary spectrum access mechanism (details in Section \ref{sec:Evolutionary-Game-Approach}). With incomplete information, users update the channel estimations based on the current access results, and make the channel selection decisions according to the distributed learning mechanism (details in Section \ref{sec:Evolutionarily-Stable-Learning}).
\end{itemize}
\end{itemize}

Suppose that $k_{m}$ users choose an idle channel $m$ to access. Then
the probability that a user $n$ (out of the $k_{m}$ users) grabs the
channel $m$ is\begin{eqnarray*}
g(k_{m}) & = & Pr\{\lambda_{n}<\min_{i\neq n}\{\lambda_{i}\}\}\\
 & = & \sum_{\lambda=1}^{\lambda_{\max}}Pr\{\lambda_{n}=\lambda\}Pr\{\lambda<\min_{i\neq n}\{\lambda_{i}\}|\lambda_{n}=\lambda\}\\
 & = & \sum_{\lambda=1}^{\lambda_{\max}}\frac{1}{\lambda_{\max}}\left(\frac{\lambda_{\max}-\lambda}{\lambda_{\max}}\right)^{k_{m}-1},\end{eqnarray*}
which is a decreasing function of the number of total contending users
$k_{m}$. Then the expected throughput of a secondary user $n$ choosing a
channel $m$ is given as\begin{equation}
U_{n}=\theta_{m}B_{m}g(k_{m}).\label{eq:u1}\end{equation}

For the ease of exposition, we will focus on the analysis of the proposed spectrum access mechanisms in the many-users
regime. Numerical results show that our algorithms also apply
when the number of users is small (see Section \ref{SUP} for the details). Since our analysis is from secondary users\textquoteright{} perspective,
we will use  terms \textquotedblleft{}secondary user\textquotedblright{}
and ``user'' interchangeably.

\section{\label{sec:Overview-of-Evolutionary}Overview of Evolutionary Game
Theory}

For the sake of completeness, we will briefly describe the background of evolutionary game theory. Detailed introduction can be found in \cite{key-10}. Evolutionary game theory was first used in biology to study the change of animal populations, and then later applied in economics to model human behaviors. It is most useful to understand how a large population of users converge to Nash equilibria in a dynamic system \cite{key-10}. A player in an evolutionary game has bounded rationality, i.e., limited computational capability and knowledge, and improves its decisions as it learns about the environment over time \cite{key-10}.

The evolutionarily stable strategy (ESS) is a key concept to describe
the evolutionary equilibrium. For simplicity, we will introduce the ESS definition (the strict Nash equilibrium in Definition \ref{def2}, respectively) in the context of a symmetric game where all users adopt the same strategy $i$ at the ESS (strict Nash equilibrium, respectively). The definition can be (and will be) extended to the case of asymmetric game \cite{key-10}, where we view the population's collective behavior as a mixed strategy $i$ at the ESS (strict Nash equilibrium, respectively).

An ESS ensures the stability such that
the population is robust to perturbations by a small fraction of players. Suppose that a small share $\epsilon$ of players in the population
deviate to choose a mutant strategy $j$, while all other players
stick to the incumbent strategy $i$. We denote the population state of the game as $\boldsymbol{x}_{(1-\epsilon)i+\epsilon j}=\left(x_{i}=1-\epsilon,x_{j}=\epsilon,x_{l}=0,\forall l\neq i,j\right)$, where $x_{a}$ denotes the fraction of users choosing strategy $a$, and the corresponding payoff of choosing strategy $a$ as $R(a,\boldsymbol{x}_{(1-\epsilon)i+\epsilon j})$.
\begin{defn}
[\!\!\cite{key-10}] \label{def1}A strategy $i$ is an \textbf{evolutionarily stable strategy}
if for every strategy $j\neq i$, there exists an $\bar{\epsilon}\in(0,1)$
such that $
R(i,\boldsymbol{x}_{\epsilon j+(1-\epsilon)i})>R(j,\boldsymbol{x}_{\epsilon j+(1-\epsilon)i})$
for any $j\neq i$ and $\epsilon\in(0,\bar{\epsilon})$.
\end{defn}

Definition \ref{def1} means that the mutant strategy $j$ cannot invade the population when the perturbation is small enough,  if the incumbent strategy $i$ is an ESS. It is shown in \cite{key-10} that any strict Nash equilibrium in noncooperative games is also an ESS.

\begin{defn}
[\!\!\cite{key-10}] \label{def2} A strategy $i$ is a \textbf{strict Nash equilibrium}
if for every strategy $j\neq i$, it satisfies that $R(i,i,...,i)>R(j,i,...,i)$,
where $R(a,i,...,i)$ denotes the payoff of choosing strategy $a\in\{i,j\}$
given other players adhering to the strategy $i$.
\end{defn}

To understand that a strict Nash is an ESS, we can set $\epsilon\rightarrow0$ in Definition
\ref{def1}, which leads to $R(i,\boldsymbol{x}_{i})>R(j,\boldsymbol{x}_{i}),\forall j\neq i$, i.e., given that almost all other players play the incumbent strategy $i$, choosing any mutant strategy $j\neq i$ will lead to a loss in payoff.

Several recent results  applied the evolutionary game theory to study various networking problems. Niyato and Hossain in \cite{key-27} investigated the evolutionary dynamics of heterogeneous network selections. Zhang \emph{et al.} in \cite{key-25} designed incentive schemes for resource-sharing  in P2P networks based on the evolutionary
game theory. Wang \emph{et al.} in \cite{key-26} proposed the evolutionary game approach for collaborative spectrum sensing mechanism design in cognitive radio networks. According to Definition \ref{def1}, the ESS obtained in \cite{key-27,key-25,key-26} is locally evolutionarily stable (i.e., the mutation $\epsilon$ is small enough). Here we apply the evolutionary game theory to design spectrum access mechanism, which can achieve global evolutionary stability (i.e., the mutation $\epsilon$ can be arbitrarily large).
%

\section{\label{sec:Evolutionary-Game-Approach}Evolutionary Spectrum Access}
We now apply the evolutionary game theory to design an efficient and stable
spectrum access mechanism with complete network information. We will show that the spectrum access equilibrium is an ESS, which guarantees
that the spectrum access mechanism is robust to random perturbations
of users' channel selections.

\begin{algorithm}[tt]
\begin{algorithmic}[1]
\State \textbf{initialization:}
\State \hspace{0.4cm} \textbf{set} the global strategy adaptation factor $\alpha\in(0,1]$.
\State \hspace{0.4cm} \textbf{select} a random channel for each user.
\State \textbf{end initialization \newline}
\Loop{ for each time slot $t$ and each user $n\in\mathcal{N}$ in parallel:}
\State \textbf{sense} and  \textbf{contend} for the chosen channel and transmit data packets if successfully grabbing the channel.
\State \textbf{broadcast} the chosen channel ID to other users through a common control channel.
\State \textbf{receive} the information of other users' channel selection and calculate the population state $\boldsymbol{x}(t)$.
\State \textbf{compute} the expected payoff $U_{n}(a_{n},\boldsymbol{x}(t))$ and the system average payoff $U(\boldsymbol{x}(t))$ according to (\ref{eq:01}) and (\ref{eq:02}), respectively.
\If{$U_{n}(a_{n},\boldsymbol{x}(t))<U(\boldsymbol{x}(t))$}
    \State \textbf{generate} a random  value $\delta$ according to a uniform distribution on $(0,1)$.
    \If{$\delta<\frac{\alpha}{x_{a_{n}}(t)}\left(1-\frac{U_{n}(a_{n},\boldsymbol{x}(t))}{U(\boldsymbol{x}(t))}\right)$}
        \State \textbf{select} a better channel $m$ with probability\[p_{m}=\frac{\max\left\{\theta_{m}B_{m}g(Nx_{m}(t))-U(\boldsymbol{x}(t)),0\right\}}{\sum_{m'=1}^{M}\max\left\{\theta_{m'}B_{m'}g(Nx_{m'}(t))-U(\boldsymbol{x}(t)),0\right\}}.\label{eq:ESA-2}\]
      \Else{ \textbf{select} the original channel.}
    \EndIf
\EndIf
\EndLoop
\end{algorithmic}
\caption{\label{alg:Evolutionary-Spectrum-Sharing}Evolutionary Spectrum Access Mechanism}
\end{algorithm}

\subsection{Evolutionary Game Formulation}

The evolutionary spectrum access game is formulated as follows:
\begin{itemize}
\item Players: the set of users $\mathcal{N}=\{1,2,...,N\}$.
\item Strategies: each user can access one of the set of  channels $\mathcal{M}=\{1,2,...,M\}$.
\item Population state: the user distribution over $M$ channels at time
$t$, $\boldsymbol{x}(t)=(x_{m}(t),\forall m \in \mathcal{M})$, where
$x_{m}(t)$ is the proportion of users selecting channel $m$ at time
$t$. We have $\sum_{m\in\mathcal{M}} x_{m}(t)=1$ for all $t$.
\item Payoff: a user $n$'s expected throughput $U_{n}(a_{n},\boldsymbol{x}(t))$
when choosing channel $a_{n}\in\mathcal{M}$, given that the population state is
$\boldsymbol{x}(t)$. Since each user has the information of channel statistics, from (\ref{eq:u1}), we have \begin{align}
U_{n}(a_{n},\boldsymbol{x}(t))=\theta_{a_{n}}B_{a_{n}}g(Nx_{a_{n}}(t)).\label{eq:01}\end{align} We also denote the system arithmetic average payoff under  population state $\boldsymbol{x}(t)$ as \begin{align} U(\boldsymbol{x}(t))=\frac{1}{M}\sum_{m=1}^{M}\theta_{m}B_{m}g(Nx_{m}(t)).\label{eq:02}\end{align}
\end{itemize}

\subsection{Evolutionary Dynamics}
Based on the evolutionary game formulation above, we propose an evolutionary
spectrum access mechanism in Algorithm \ref{alg:Evolutionary-Spectrum-Sharing} by reversing-engineering the replicator dynamics. The idea is to let those users who have  payoffs lower than the system average payoff $U(\boldsymbol{x}(t)$ to select a better channel, with a probability proportional to the (normalized) channel's \textquotedblleft{}net fitness\textquotedblright{} $\theta_{m}B_{m}g(Nx_{m}(t))-U(\boldsymbol{x}(t))$.
We show
that the dynamics of channel selections in the mechanism can be described with the evolutionary
dynamics in (\ref{eq:ESS-1}). The proof is given in the Appendix.

\begin{thm}\label{thm:For-the-evolutionary-1}
For the evolutionary spectrum access mechanism in Algorithm \ref{alg:Evolutionary-Spectrum-Sharing}, the evolutionary dynamics are given as
\begin{align}
\dot{x}_{m}(t)=\alpha\left(\frac{U_{n}(m,\boldsymbol{x}(t))}{U(\boldsymbol{x}(t))}-1\right),\forall m\in\mathcal{M},\label{eq:ESS-1}\end{align}
where the derivative is with respect to time $t$. \end{thm}

\subsection{\label{AESS}Evolutionary Equilibrium in Asymptotic Case $\lambda_{max}=\infty$}
We next investigate the equilibrium of the  evolutionary
spectrum access mechanism. To obtain useful insights, we first focus on the asymptotic case where the number of backoff mini-slots $\lambda_{max}$
goes to $\infty$, such that \begin{eqnarray}
g(k) & = & \lim_{\lambda_{max}\rightarrow\infty}\sum_{\lambda=1}^{\lambda_{max}}\frac{1}{\lambda_{max}}\left(\frac{\lambda_{max}-\lambda}{\lambda_{max}}\right)^{k-1} \nonumber \\
 & = & \lim_{\frac{1}{\lambda_{max}}\rightarrow0}\sum_{\lambda=0}^{\lambda_{max}-1}\left(\frac{\lambda}{\lambda_{max}}\right)^{k-1}\frac{1}{\lambda_{max}} \nonumber \\
 & = & \int_{0}^{1}z^{k-1}dz=\frac{1}{k}.\end{eqnarray}

This is a good approximation when the number of mini-slots $\lambda_{max}$ for backoff is much larger than the
number of users $N$ and collisions rarely occur. In this case, $U_{n}(a_{n},\boldsymbol{x}(t))=\frac{\theta_{a_{n}}B_{a_{n}}}{Nx_{m}(t)}$ and $U(\boldsymbol{x}(t))=\frac{\sum_{i=1}^{M}\theta_{i}B_{i}}{N}$. According to Theorem \ref{thm:For-the-evolutionary-1}, the evolutionary dynamics in (\ref{eq:ESS-1}) become\begin{align}\dot{x}_{m}(t)=\alpha\left(\frac{\frac{\theta_{m}B_{m}}{x_{m}(t)}}{\frac{1}{M}\sum_{i=1}^{M}\frac{\theta_{i}B_{i}}{x_{i}(t)}}-1\right).\label{eq:12}\end{align}
From (\ref{eq:12}), we have
\begin{thm}\label{thm22}
The evolutionary spectrum access mechanism in asymptotic case $\lambda_{max}=\infty$ globally converges to the evolutionary
equilibrium $\boldsymbol{x}^{*}=\left(x_{m}^{*}=\frac{\theta_{m}B_{m}}{\sum_{i=1}^{M}\theta_{i}B_{i}},\forall m\in\mathcal{M}\right)$.
\end{thm}

The proof is given in the Appendix. Theorem \ref{thm22} implies that
\begin{cor}
The evolutionary spectrum access mechanism converges to the equilibrium
$\boldsymbol{x}^{*}$ such that users on different channels achieve
the same expected throughput, i.e.,\begin{align}
U_{n}(m,\boldsymbol{x}^{*})=U_{n}(m',\boldsymbol{x}^{*}),\forall m,m'\in\mathcal{M}.\label{eq:aess} \end{align}
\end{cor}

We next show that for the general case $\lambda_{max}<\infty$, the evolutionary dynamics also globally converges to the ESS equilibrium as given in (\ref{eq:aess}).

\subsection{Evolutionary Equilibrium in General Case $\lambda_{max}<\infty$}
For the general case $\lambda_{max}$, since the channel grabbing probability $g(k)$ does not have the close-form expression, it is hence difficult to obtain the equilibrium solution of differential equations in (\ref{eq:ESS-1}). However, it is easy to verify that the equilibrium $\boldsymbol{x}^{*}$ in (\ref{eq:aess}) is also a stationary point such that the evolutionary dynamics (\ref{eq:ESS-1}) in the general case $\lambda_{max}<\infty$ satisfy $\dot{x}_{m}(t)=0.$ Thus, at the equilibrium $\boldsymbol{x}^{*}$, users on different channels achieve the same expected throughput.

We now study the evolutionary stability of the equilibrium. In general, the equilibrium of the replicator dynamics may not be
an ESS \cite{key-10}. For our model, we can prove the following.
\begin{thm}
\label{thm:For-the-evolutionary}For the evolutionary spectrum access
mechanism, the evolutionary equilibrium $\boldsymbol{x}^{*}$ in (\ref{eq:aess})
is an ESS. \end{thm}
The proof is given in Section \ref{proof3}. Actually we can obtain a stronger result than Theorem \ref{thm:For-the-evolutionary}.
Typically, an ESS is only locally asymptotically
stable (i.e., stable within a limited region around the ESS) \cite{key-10}.  For our case, we show that
the evolutionary equilibrium $\boldsymbol{x}^{*}$ is globally asymptotically stable (i.e., stable in the entire feasible region of a population
state $\boldsymbol{x}$, $\{\boldsymbol{x}=(x_{m},m\in\mathcal{M})|\sum_{m=1}^{M}x_{m}=1\ \mbox{and}\ x_{m}\geq0,\forall m\in\mathcal{M}\}$).

To proceed, we first define the following function \begin{equation}
L(\boldsymbol{x})=\sum_{m=1}^{M}\int_{-\infty}^{x_{m}}\theta_{m}B_{m}g(Nz)dz.\label{eq:L1}\end{equation}
Since $g(\cdot)$ is a decreasing function, it is easy to check that
the Hessian matrix of $L(\boldsymbol{x})$ is negative definite. It
follows that $L(\boldsymbol{x})$ is strictly concave and hence has
a unique global maximum $L^{*}$. By the first order condition, we obtain the
optimal solution $\boldsymbol{x}^{*}$, which is the same as the evolutionary equilibrium $\boldsymbol{x}^{*}$ in (\ref{eq:aess}). Then by showing that $V(\boldsymbol{x}(t))=L^{*}-L(\boldsymbol{x}(t))$ is a strict Lyapunov function, we have

\begin{thm}
\label{thm:For-the-replicator2}For the evolutionary spectrum access
mechanism, the evolutionary equilibrium $\boldsymbol{x}^{*}$ in (\ref{eq:aess}) is globally asymptotically stable.\end{thm}

The proof is given in the Appendix. Since the ESS is globally asymptotically stable, the evolutionary spectrum access mechanism is robust to any degree of (not necessarily small) random perturbations of channel selections. 
\section{\label{sec:Evolutionarily-Stable-Learning}Learning Mechanism For
Distributed spectrum access}

For the evolutionary spectrum access mechanism in Section \ref{sec:Evolutionary-Game-Approach}, we assume that
each user has the perfect knowledge of channel statistics and the population state by information exchange on a common control channel. Such mechanism leads to significant communication overhead and energy consumption, and may even be impossible in some systems. We thus propose
a learning mechanism for distributed spectrum access with incomplete information. The challenge is how to achieve the evolutionarily stable state based on user's local observations only.

\begin{algorithm}[tt]
\begin{algorithmic}[1]
\State \textbf{initialization:}
\State \hspace{0.4cm} \textbf{set} the global memory weight $\gamma\in(0,1)$ and the set of accessed channels $\mathcal{M}_{n}=\varnothing$ for each user $n$.
\State \textbf{end initialization\newline}

\Loop{ for each user $n\in\mathcal{N}$ in parallel:\newline\newline\Comment{\textbf{\emph{Initial Channel Estimation Stage}}}}

    \While{$\mathcal{M}_{n}\neq\mathcal{M}$}
        \State \textbf{choose} a channel $m$ from the set $\mathcal{M}_{n}^{c}$ randomly.
        \State \textbf{sense} and \textbf{contend} to access the channel $m$ at each time slot of the decision period.
        \State \textbf{estimate} the expected throughput $\tilde{U}_{m,n}(0)$ by (\ref{bb}).
        \State \textbf{set} $\mathcal{M}_{n}=\mathcal{M}_{n}\cup\{m\}$.
    \EndWhile{\newline\newline\Comment{\textbf{\emph{Access Strategy Learning Stage}}}}

    \For{ for each time period $T$}
        \State \textbf{choose} a channel $m$ to access according to the mixed strategy $\boldsymbol{f}_{n}(T)$  in (\ref{eq:ESL-2}).
        \State \textbf{sense} and \textbf{contend} to access the channel $m$ at each time slot of the decision period.
        \State \textbf{estimate}  the qualities of the chosen channel $m$ and the unchosen channels $m'\neq m$ by (\ref{est1}) and (\ref{est2}), respectively.
    \EndFor
\EndLoop

\end{algorithmic}
\caption{\label{alg:Learning}Learning Mechanism For Distributed Spectrum Access}
\end{algorithm}

\subsection{Learning Mechanism For Distributed Spectrum Access}
\begin{figure}[tt]
\centering
\includegraphics[scale=0.6]{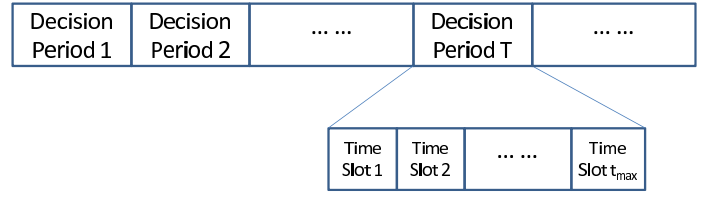}
\caption{\label{fig:Decision}Learning time structure}
\end{figure}

The proposed learning process is shown in Algorithm \ref{alg:Learning} and has two sequential stages: \emph{initial channel estimation} (line $5$ to $10$)
and \emph{access strategy learning} (line $11$ to $15$). Each stage is defined over a sequence of decision periods $T=1,2,...$, where each decision period consists of $t_{\max}$ time slots (see Figure \ref{fig:Decision} as an illustration).

The key idea of distributed learning here is to adapt each user's spectrum access decision based on its accumulated experiences. In the first stage, each user initially estimates the expected throughput
by accessing all the channels in a randomized round-robin manner. This ensures that all users do not choose the same channel at the same period. Let $\mathcal{M}_{n}$ (equals to $\varnothing$ initially) be the set of channels accessed by user $n$
 and $\mathcal{M}_{n}^{c}=\mathcal{M}\backslash\mathcal{M}_{n}$. At
beginning of each decision period, user $n$ randomly chooses
a channel $m\in\mathcal{M}_{n}^{c}$ (i.e., a channel that has not been accessed before) to access. At end of the period,
user $n$ can estimate the expected throughput by sample averaging as\begin{equation}
Z_{m,n}(0)=(1-\gamma)\frac{\sum_{t=1}^{t_{\max}}b_{m}(t)I_{\{a_{n}(t,T)=m\}}}{t_{\max}},\label{bb}\end{equation}
where $0<\gamma<1$ is called the memory weight and $I_{\{a_{n}(t,T)=m\}}$ is an indicator function and equals $1$ if the channel $m$ is idle  at time slot $t$ and
the user $n$ chooses and successfully grabs the channel $m$. Motivation of multiplying $(1-\gamma)$ in (\ref{bb}) is to scale down the impact of the noisy instantaneous estimation on the learning.  Note that there are $t_{\max}$ time slots within each decision period, and thus the user will be able to have a fairly good estimation of the expected throughput if $t_{\max}$ is reasonably large. Then user $n$ updates the set of accessed channels as $\mathcal{M}_{n}=\mathcal{M}_{n}\cup\{m\}$. When all the channels are accessed, i.e., $\mathcal{M}_{n}=\mathcal{M}$, the stage
of initial channel estimation ends. Thus, the total time slots for the first stage is $Mt_{\max}$.

In the second stage, at each period $T\geq 1$, each user $n\in\mathcal{N}$ selects a
channel $m$ to access according to a mixed strategy
$\boldsymbol{f}_{n}(T)=(f_{1,n}(T),...,f_{M,n}(T))$, where $f_{m,n}(T)$
is the probability of user $n$ choosing channel $m$ and is computed as\begin{equation}
f_{m,n}(T)=\frac{\sum_{\tau=0}^{T-1}\gamma^{T-\tau-1}Z_{m,n}(\tau)}{\sum_{i=1}^{M}\sum_{\tau=0}^{T-1}\gamma^{T-\tau-1}Z_{i,n}(\tau)},\forall m\in\mathcal{M}.\label{eq:ESL-2}\end{equation}
Here $Z_{m,n}(\tau)$
is user $n$'s estimation of the quality of channel $m$ at period $\tau$ (see (\ref{est2}) and(\ref{est1}) later). The update in (\ref{eq:ESL-2}) means
that each user adjusts its mixed strategy according to its weighted average estimations of
all channels' qualities.

Suppose that user $n$ chooses channel $m$ to access at period $\tau$. For the unchosen channels $m'\neq m$ at this period, user $n$ can empirically estimate the quality of this channel according to its past memories as \begin{equation}
Z_{m',n}(\tau)= (1-\gamma)\sum_{\tau^{'}=0}^{\tau-1}\gamma^{\tau-\tau^{'}-1}Z_{m',n}(\tau^{'}).\label{est2}\end{equation} For the chosen channel $m$, user $n$ will update the estimation of this channel $m$ by combining the empirical estimation with the real-time throughput measurement in this period, i.e., \begin{align}
Z_{m,n}(\tau)  = & (1-\gamma)\left(\sum_{\tau^{'}=0}^{\tau-1}\gamma^{\tau-\tau^{'}-1}Z_{m,n}(\tau^{'})\right.\nonumber \\
& \left. + \frac{\sum_{t=1}^{t_{\max}}b_{m}(t)I_{\{a_{n}(t,\tau)=m\}}}{t_{\max}}\right).\label{est1}\end{align}

\subsection{Convergence of Learning Mechanism}
We now study the convergence of the learning mechanism. Since each user only utilizes its local estimation to adjust its mixed channel access strategy,  the exact ESS is difficult to achieve due to the random estimation noise. We will show that the learning mechanism can converge to the ESS on time average.

According to the theory of stochastic approximation \cite{kushner1984approximation}, the limiting behaviors of the learning mechanism with the random estimation noise can be well approximated by the corresponding mean dynamics. We thus study the mean dynamics of the learning mechanism. To proceed, we define the mapping from the mixed channel access strategies $\boldsymbol{f}(T)=(\boldsymbol{f}_{1}(T),...,\boldsymbol{f}_{N}(T))$
to the mean throughput of user $n$ choosing channel $m$ as $Q_{m,n}(\boldsymbol{f}(T))\triangleq E[U_{n}(m,\boldsymbol{x}(T))|\boldsymbol{f}(T)]$.
Here the expectation $E[\cdot]$ is taken with respective to the mixed
strategies $\boldsymbol{f}(T)$ of all users. We show that
\begin{thm}\label{lm2}
As the memory weight $\gamma\rightarrow1$, the mean dynamics of the learning mechanism for distributed spectrum access are given as ($\forall m\in\mathcal{M},n\in\mathcal{N}$)\begin{align}
\dot{f}_{m,n}(T)=f_{m,n}(T)\left(Q_{m,n}(\boldsymbol{f}(T))-\sum_{i=1}^{M}f_{i,n}(T)Q_{i,n}(\boldsymbol{f}(T))\right),\label{learnD}\end{align}
where the derivative is with respect to period $T$.\end{thm}
The proof is given in Section \ref{proof2}. Interestingly, similarly with the evolutionary dynamics in (\ref{eq:ESS-1}), the learning dynamics in (\ref{learnD}) imply
that if a channel offers a higher throughput for a user than the user's average throughput over all channels, then the user will exploit that channel more often in the future learning. However, the evolutionary dynamics in (\ref{eq:ESS-1}) are based on the population level with complete network information, while the learning dynamics in (\ref{learnD}) are derived from  the individual local estimations. We show in Theorem \ref{thm:haha} that the mean dynamics of learning mechanism converge to the ESS in (\ref{eq:aess}), i.e., $Q_{m,n}(\boldsymbol{f}^{*})=Q_{m',n}(\boldsymbol{f}^{*})$.

\begin{thm} \label{thm:haha}
As the memory weight $\gamma\rightarrow1$, the mean dynamics of the learning mechanism
for distributed spectrum access asymptotically converge to a limiting
point $\boldsymbol{f}^{*}$ such that \begin{align}
Q_{m,n}(\boldsymbol{f}^{*})=Q_{m',n}(\boldsymbol{f}^{*}),\forall m,m'\in\mathcal{M},\forall n\in\mathcal{N}.\label{eq:qq}\end{align}
\end{thm}

The proof is given in Section \ref{proof6}. \rev{Since $Q_{m,n}(\boldsymbol{f}^{*})= E[U_{n}(m,\boldsymbol{x}(T))|\boldsymbol{f}^{*}]$ and the mean dynamics converge to the equilibrium $\boldsymbol{f}^{*}$ satisfying (\ref{eq:qq}) (i.e., $E[U_{n}(m,\boldsymbol{x}(T))|\boldsymbol{f}^{*}]=E[U_{n}(m',\boldsymbol{x}(T))|\boldsymbol{f}^{*}]$),  the learning mechanism thus converges to the ESS (\ref{eq:aess}) (achieved by the evolutionary spectrum access mechanism) on the time average. Note that both the evolutionary spectrum access mechanism in Algorithm \ref{alg:Evolutionary-Spectrum-Sharing} and learning mechanism in Algorithm \ref{alg:Learning} involve basic arithmetic operations and random number generation over $M$ channels, and hence have a linear computational complexity of $\mathcal{O}(M)$ for each iteration. However, due to the incomplete information, the learning mechanism typically takes a longer convergence time in order to get a good estimation of the environment.}

\section{\label{sec:Simulation-Results}Simulation Results}
In this section, we evaluate the proposed algorithms by simulations. We consider a cognitive radio network consisting
$M=5$ Rayleigh fading channels. The channel idle probabilities are $\{\theta_{m}\}_{m=1}^{M}=\{\frac{2}{3},\frac{4}{7},\mbox{\ensuremath{\frac{5}{9}},\ensuremath{\frac{1}{2},\frac{4}{5}}}\}$. The data rate on a channel $m$ is computed according to the Shannon capacity, i.e., $b_{m}=\zeta_{m}\log_{2}(1+\frac{P_{n}h_{m}}{N_{0}})$,
where $\zeta_{m}$ is the bandwidth of channel $m$, $P_{n}$ is the
power adopted by users, $N_{0}$ is the noise power, and
$h_{m}$ is the channel gain (a realization of a random variable that
follows the exponential distribution with the mean $\bar{h}_{m}$).
In the following simulations, we set $\zeta_{m}=10$ MHz, $N_{0}=-100$
dBm, and $P_{n}=100$ mW. By choosing different mean channel gain $\bar{h}_{m}$, we have different mean data rates $B_{m}=E[b_{m}]$, which equal $15, 70, 90, 20$ and $100$ Mbps, respectively.

\begin{figure}
\centering
\includegraphics[scale=0.53]{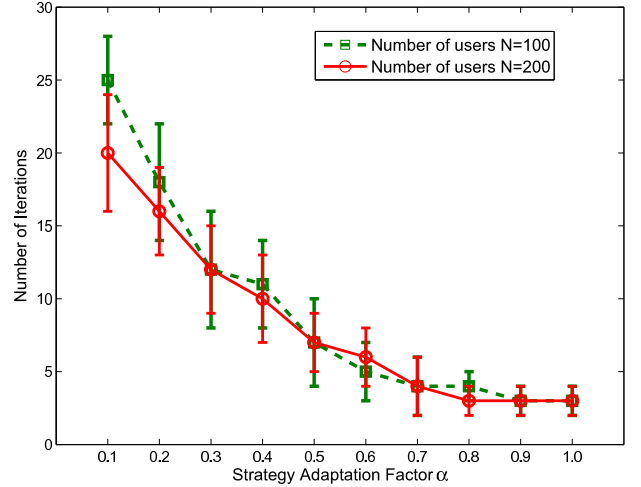}
\caption{\label{fig:Iterations}The iterations need for the convergence of the evolutionary spectrum accessing mechanism
with different choices of strategy adaptation factor $\alpha$. The confidence interval is $95\%$.}
\end{figure}

\begin{figure}
\centering
\includegraphics[scale=0.64]{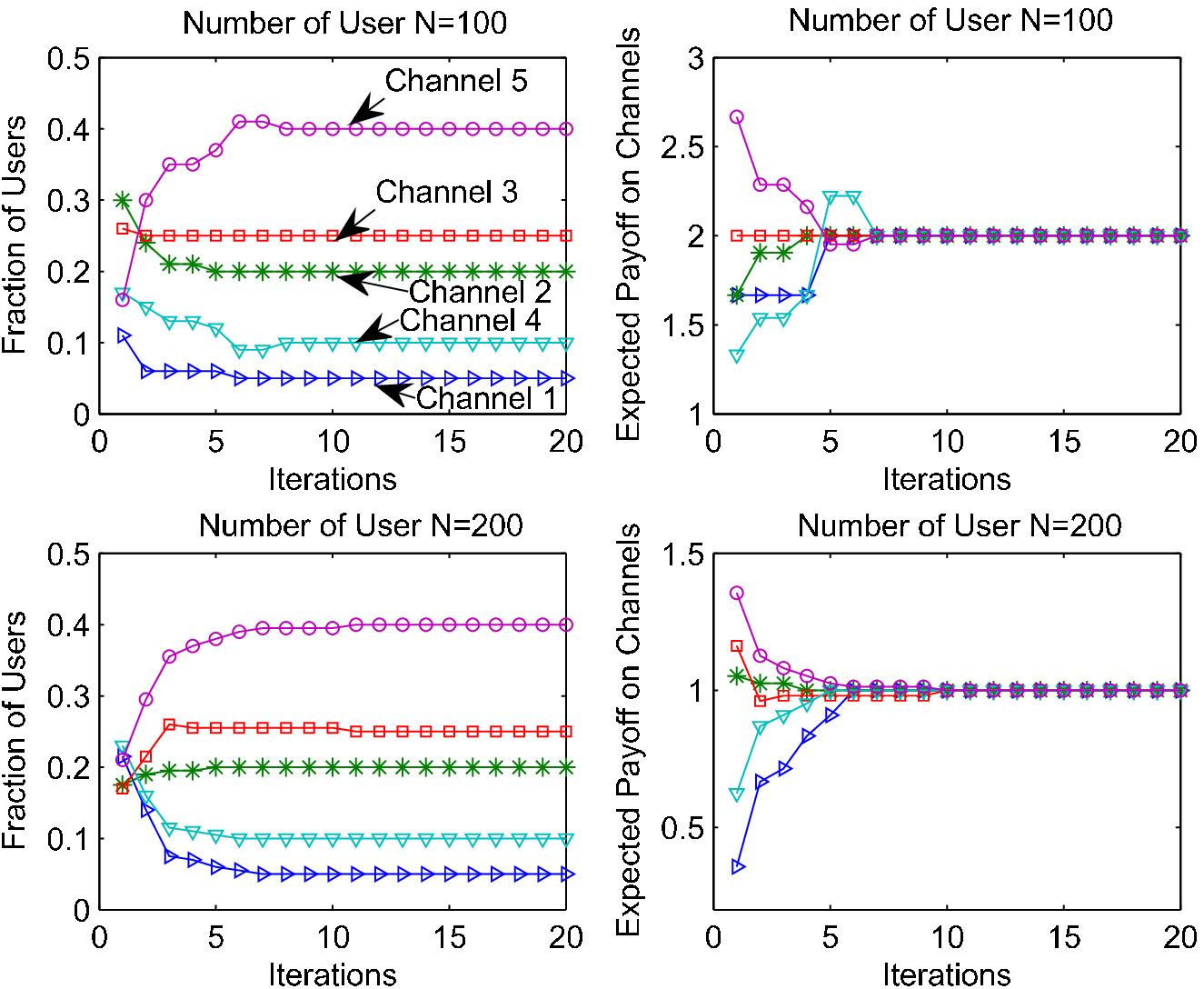}
\caption{\label{fig:The-expected-payoff}The fraction of users on each channel and the expected user payoff of accessing different
channels with the number of users $N=100$ and $200$, respectively, and the number of backoff mini-slots $\lambda_{max}=100000$.}
\end{figure}

\begin{figure}
\centering
\includegraphics[scale=0.64]{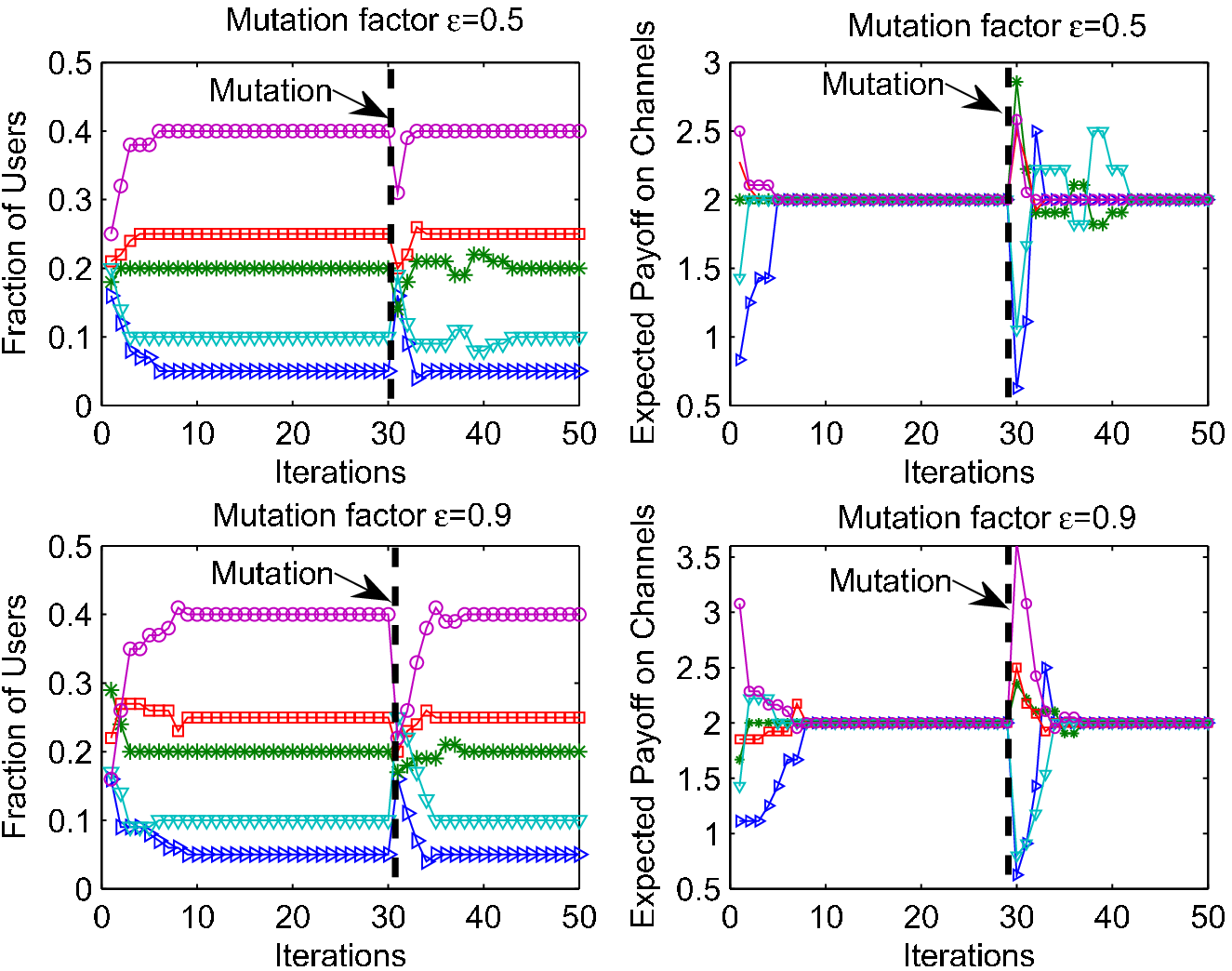}
\caption{\label{fig:Stability-of-the}Stability of the evolutionary spectrum
access mechanism. Fraction of users in total $N=200$ users who
choose mutant channels randomly at time slot $30$ equal to $0.5$ and $0.9$, respectively, and the number of backoff mini-slots $\lambda_{max}=100000$.}
\end{figure}

\begin{figure}
\centering
\includegraphics[scale=0.63]{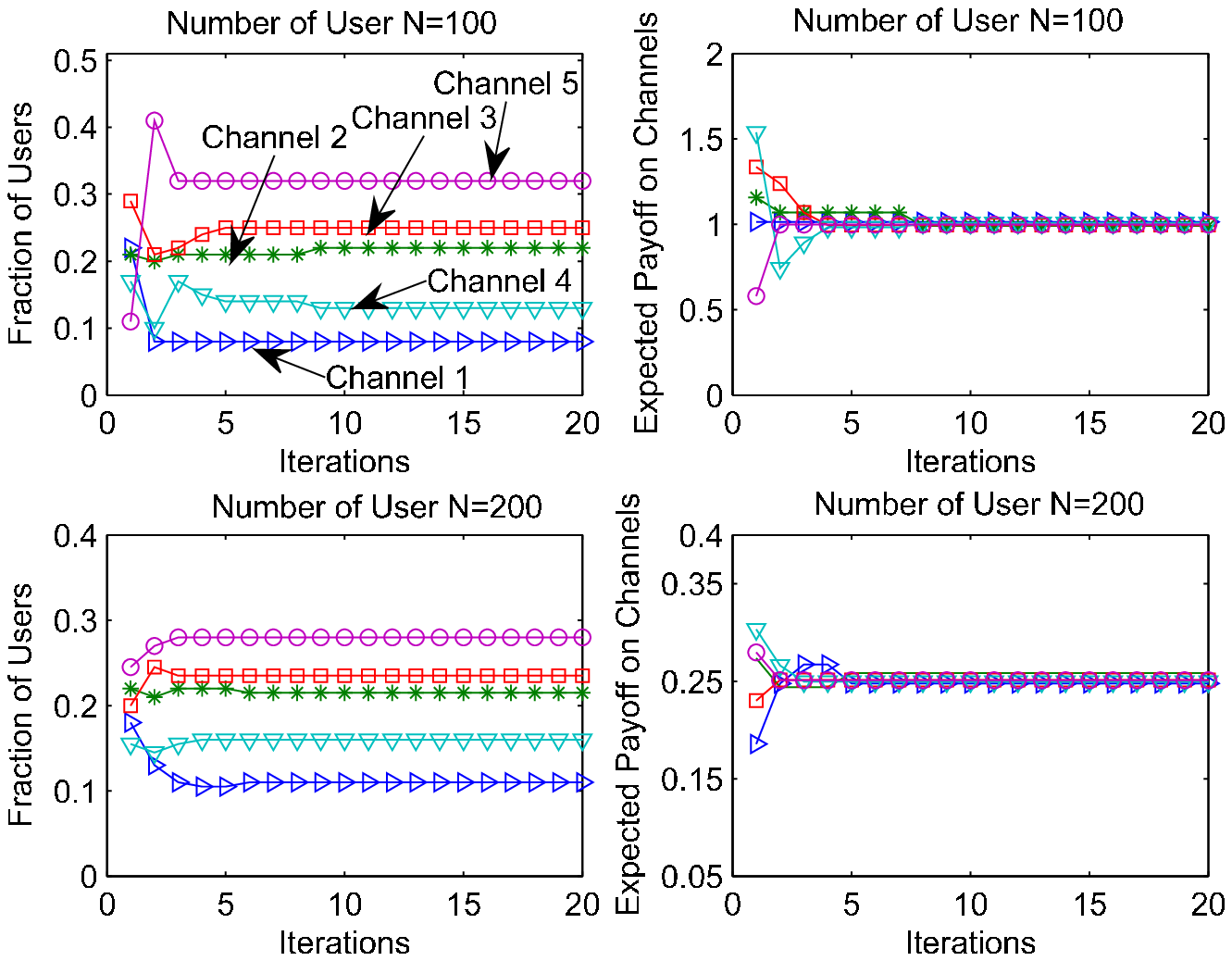}
\caption{\label{fig:The-expected-payoff1}The fraction of users on each channel  and the expected user payoff of accessing different
channels with the number of users $N=100$ and $200$, respectively, and the number of backoff mini-slots $\lambda_{max}=20$.}
\end{figure}

\begin{figure}
\centering
\includegraphics[scale=0.63]{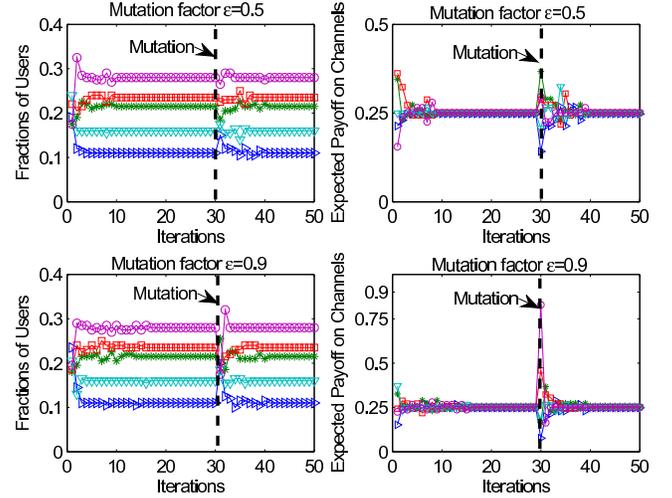}
\caption{\label{fig:Stability-of-the1}Stability of the evolutionary spectrum
access mechanism. Fraction of users in total $N=200$ users who
choose mutant channels randomly at time slot $30$ equal to $0.5$ and $0.9$, respectively, and the number of backoff mini-slots $\lambda_{max}=20$.}
\end{figure}

\subsection{Evolutionary Spectrum Access in Large User Population Case}
We first study the evolutionary spectrum mechanism with complete network information in Section \ref{sec:Evolutionary-Game-Approach} with a large user population. We found
that the convergence speed of the evolutionary spectrum access mechanism increases as the strategy adaptation factor $\alpha$ increases (see Figure \ref{fig:Iterations}). We set the strategy adaptation
factor $\alpha=0.5$ in the following simulations in order to better demonstrate the evolutionary dynamics. We implement the
evolutionary spectrum access mechanism with the number of users
$N=100$ and $200$, respectively, in both large and small $\lambda_{max}$ cases.

\subsubsection{Large $\lambda_{max}$ Case}
We first consider the case that the number of backoff mini-slots $\lambda_{max}=100000$, which is much larger that the number of users $N$ and thus collisions in channel contention rarely occur. This case can be approximated by the asymptotic case $\lambda_{max}=\infty$ in Section \ref{AESS}. The simulation results are shown in Figures
\ref{fig:The-expected-payoff} and \ref{fig:Stability-of-the}.
From these figures, we see that
\begin{itemize}
\item \emph{Fast convergence: }the algorithm takes less than $20$ iterations
to converge in all cases (see Figure \ref{fig:The-expected-payoff}).
\item \emph{Convergence to ESS}: in both $N=100$ and $200$ cases, the algorithm converges
to the ESS $\boldsymbol{x}^{*}=\left(\frac{\theta_{1}B_{1}}{\sum_{i=1}^{M}\theta_{i}B_{i}},...,\frac{\theta_{M}B_{M}}{\sum_{i=1}^{M}\theta_{i}B_{i}}\right)$
(see Figure the left column of \ref{fig:The-expected-payoff}).
At the ESS $\boldsymbol{x}^{*}$,
each user achieves the same expected payoff $U_{n}(a_{n}^{*},\boldsymbol{x}^{*})=\frac{\sum_{i=1}^{M}\theta_{i}B_{i}}{N}$
(see the right column of Figure \ref{fig:The-expected-payoff}).
\item \emph{Asymptotic stability}: to investigate the stability of the evolutionary
spectrum access mechanism, we let a fraction of users play the
mutant strategies when the system is at the ESS $\boldsymbol{x}^{*}$.
At time slot $t=30$, $\epsilon=0.5$ and $0.9$ fraction of
users will randomly choose a new channel. The result is shown in Figure
$\ref{fig:Stability-of-the}$. We see that the algorithm is capable
to recover the ESS $\boldsymbol{x}^{*}$ quickly after the mutation
occurs. This demonstrates that the evolutionary spectrum access
mechanism is robust to the perturbations in the network.
\end{itemize}

\subsubsection{Small  $\lambda_{max}$ Case}
We now consider the case that the number of backoff mini-slots $\lambda_{max}=20$, which is smaller than the number of users $N$. In this case, severe collisions in channel contention may occur and hence lead to a reduction in data rates  for all users. The results are shown in Figures \ref{fig:The-expected-payoff1} and \ref{fig:Stability-of-the1}. We see that  a small $\lambda_{max}$ leads to a system performance loss (i.e., $\sum_{n=1}^{N}U_{n}(a_{n}(T),\boldsymbol{x}(T))<\sum_{m=1}^{M}\theta_{m}B_{m}$), due to severe collisions in channel contention. However, the evolutionary spectrum access mechanism still quickly converges to the ESS as given in (\ref{eq:aess}) such that all users achieve the same expected throughput, and the asymptotic stable property also holds. This verifies the efficiency of the mechanism in the small $\lambda_{max}$ case.

\subsection{Distributed Learning Mechanism in Large User Population Case}
\begin{figure}
\centering
\includegraphics[scale=0.65]{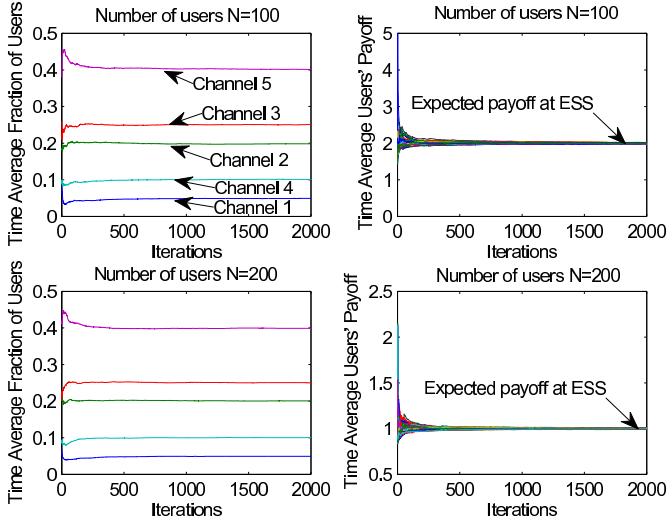}\caption{\label{fig:Learning-mechanism-for}Learning mechanism for distributed spectrum
access with the number of users $N=100$ and $200$, respectively, and the number of backoff mini-slots $\lambda_{max}=100000$.}
\end{figure}

\begin{figure}
\centering
\includegraphics[scale=0.65]{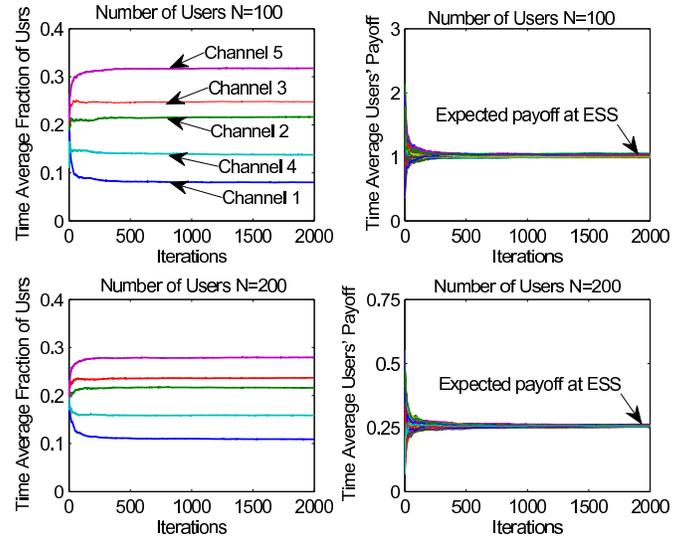}\caption{\label{fig:Learning-mechanism-for1}Learning mechanism for distributed spectrum
access with the number of users $N=100$ and $200$, respectively, and the number of backoff mini-slots $\lambda_{max}=20$.}
\end{figure}

We next evaluate the learning mechanism for distributed spectrum
access with a large user population. We implement the learning mechanism with
the number of users $N=100$ and $N=200$, respectively, in both large and small $\lambda_{max}$ cases. We set the memory factor $\gamma=0.99$ and the
length of a decision period $t_{\max}=100$ time slots, which provides a good estimation of the mean data rate.  Figures \ref{fig:Learning-mechanism-for} and \ref{fig:Learning-mechanism-for1} show the time
average user distribution on the channels converges to the ESS,
and the time average user's payoff converges the expected payoff at the ESS.  Note that users achieve this result without prior knowledge of the statistics of the channels, and the number of users utilizing each channel keeps changing in the learning scheme.

\subsection{\label{SUP}Evolutionary Spectrum Access and Distributed Learning in Small User Population Case}
\begin{figure}
\centering
\includegraphics[scale=0.55]{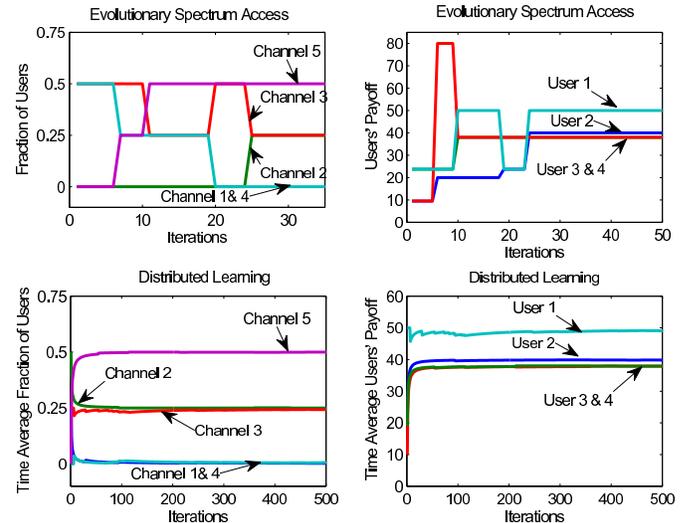}\caption{\label{fig:small}Evolutionary spectrum access and Learning mechanism for distributed spectrum
access with the number of users $N=4$, and the number of backoff mini-slots $\lambda_{max}=20$.}
\end{figure}

We then consider the case that the user population $N$ is small.
We implement the proposed evolutionary spectrum access mechanism and
distributed learning mechanism with the number of users $N=4$ and
the number of backoff mini-slots $\lambda_{\max}=20$. The results
are shown in Figure \ref{fig:small}. We see that the evolutionary spectrum access
mechanism converges to the equilibrium such that channel $5$ has
$2$ users and both channel $1$ and $2$ have $1$ user. These $4$
users achieve the expected throughput equal to $50$, $40$, $38$
and $38$ Mbps, respectively, at the equilibrium. It is easy to check
that any user unilaterally changes its channel
selection at the equilibrium  will lead to a loss in throughput, hence the equilibrium is
a strict Nash equilibrium. According to \cite{key-10},
any strict Nash equilibrium is also an ESS and hence the convergent
equilibrium is an ESS. For the distributed learning mechanism, we
see that the mechanism also converges to the same equilibrium on the
time average. This verifies that effectiveness of the proposed mechanisms
in the small user population case.

\subsection{Performance Comparison}
To benchmark the performance of the proposed mechanisms, we compare
them with the following two algorithms:
\begin{itemize}
\item \emph{Centralized optimization}: we solve the centralized optimization problem
$\max_{\boldsymbol{x}}\sum_{n=1}^{N}U_{n}(a_{n},\boldsymbol{x})$,
i.e., find the optimal population state $\boldsymbol{x}_{opt}$ that
maximizes the system throughput.
\item \emph{Distributed reinforcement learning}: we also implement the distributed
algorithm in \cite{ikey-9} by generalizing the single-agent reinforcement learning to
the multi-agent setting. More specifically, each user $n$ maintains a
perception value $P_{m}^{n}(T)$ to describe the performance of channel $m$, and select the channel $m$ with the probability $f_{m,n}(T)=\frac{e^{\nu P_{m}^{n}(T)}}{\sum_{m'=1}^{M}e^{\nu P_{m'}^{n}(T)}}$
where $\mbox{\ensuremath{\nu}}$ is called the temperature. Once a
payoff $U_{n}(T)$ is received, user $n$ updates the perception value
as $P_{m}^{n}(T+1)=(1-\mu_{T})P_{m}^{n}(T)+\mu_{T}U_{n}(T)I_{\{a_{n}(T)=m\}}$
where $\mu_{T}$ is the smooth factor satisfying $\sum_{T=1}^{\infty}\mu_{T}=\infty$
and $\sum_{T=1}^{\infty}\mu_{T}^{2}<\infty$. As shown in \cite{ikey-9},
when $\nu$ is sufficiently large, the algorithm converges to a stationary
point. We hence set $\mu_{T}=\frac{100}{T}$ and $\nu=10$ in the
simulation, which guarantees the convergence and achieves a good system performance.
\end{itemize}

\begin{figure}
\centering
\includegraphics[scale=0.53]{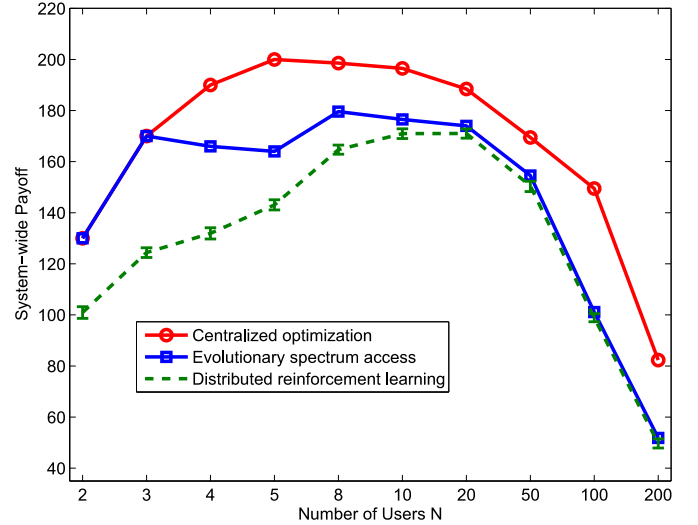}
\caption{\label{fig:Comparison-of-the}Comparison of the evolutionary spectrum
access mechanism with the distributed reinforcement learning and centralized
optimization. The confidence interval is $95\%$.}
\end{figure}

Since the proposed learning mechanism in this paper can converge to
the same equilibrium as the evolutionary spectrum access mechanism,
we only implement the evolutionary spectrum access mechanism in this
experiment. The results are shown in Figure \ref{fig:Comparison-of-the}. Since the global optimum by centralized optimization and the ESS by evolutionary spectrum access are deterministic, only the confidence interval of the distributed reinforcement learning is shown here. We see that the evolutionary spectrum access mechanism achieves up
to $35\%$ performance improvement over the distributed reinforcement
learning algorithm. Compared with the centralized optimization approach,
the performance loss of the evolutionary spectrum access mechanism is at most $38\%$. When the number of users $N$ is small (e.g., $N\leq50$), the performance loss can be further reduced to less than $25\%$. Note that the solution by the centralized optimization is not incentive compatible, since it is
not a Nash equilibrium and user can improve its payoff by changing
its channel selection unilaterally. While the evolutionary spectrum access mechanism
achieves an ESS, which is also a (strict) Nash equilibrium and evolutionarily
stable. Interestingly, the curve of the evolutionary spectrum access mechanism in Figure \ref{fig:Comparison-of-the} achieves a local minimum when the number of users $N=5$. This can be interpreted by the property of the Nash equilibrium. When the number of users $N=4$, these four users will utilize the three channels with high data rate (i.e., Channel $2$, $3$, and $5$ in the simulation).  When the number of users $N=5$, the same three channels are utilized at the Nash equilibrium. In this case, there will be a system performance loss due to severer channel contention. However, no user at the equilibrium is willing to switch to another vacant channel, since the remaining vacant channels have low data rates and such a switch will incurs a loss to the user. When the number of users $N=8$, all given channels are utilized at the Nash equilibrium, and this improves the system performance.

\begin{figure}
\centering
\includegraphics[scale=0.8]{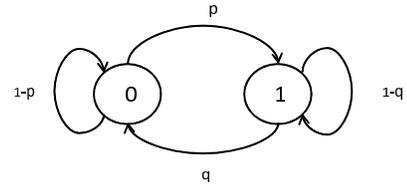}
\caption{\label{fig:Two-states-Markovian}Two states Markovian channel model}
\end{figure}

\subsection{Distributed Learning Mechanism In Markovian Channel Environment}

For the ease of exposition, we have considered the i.i.d. channel model as far. We now consider the proposed mechanisms in the Markovian
channel environment. Since in the evolutionary spectrum
access mechanism each user has the complete information aprior (including
the stationary distribution that a channel is idle), the Markovian
setting will not affect the evolutionary spectrum access mechanism.
We hence focus on evaluating the learning mechanism.

We consider a network of $N=100$ users and $M=10$ channels. The
states of channels change according to independent Markovian
processes (see Figure \ref{fig:Two-states-Markovian}). We denote
the channel state probability vector of channel $m$ at time slot
$t$ as $\boldsymbol{p}_{m}(t)\triangleq(Pr\{S_{m}(t)=0,Pr\{S_{m}(t)=1\}\}),$
which follows a two state Markov chain as $\boldsymbol{p}_{m}(t)=\boldsymbol{p}_{m}(t-1)\Gamma,\forall t\geq1$,
with the transition matrix\[
\Gamma=\left[\begin{array}{cc}
1-p & p\\
q & 1-q\end{array}\right].\]
For the simulation, we set $p=q=\epsilon,$ where $\epsilon$ is called
the dynamic factor. A larger $\epsilon$ means that the channel state
changes faster over time. The mean data rates $B_{m}$ of $10$ channels
are $10,40,50,20,80,60,15,25,30,$ and $70$ Mbps, respectively.

\begin{figure}
\centering
\includegraphics[scale=0.53]{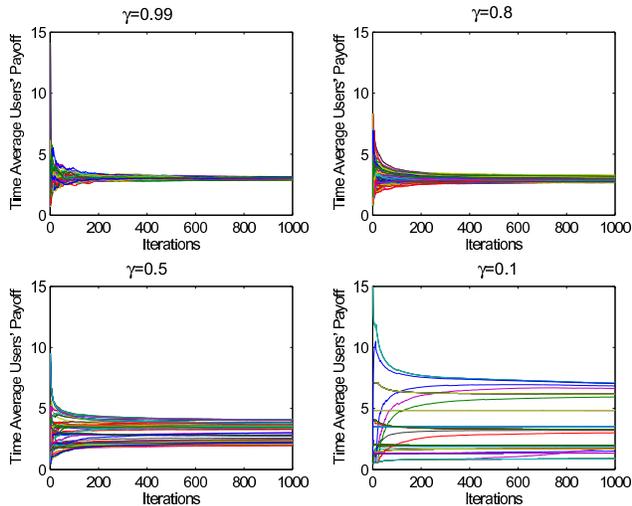}
\caption{\label{fig:Distributed-learning-mechanism}Distributed learning mechanism
with different memory weights $\gamma$}
\end{figure}

We first set the dynamic factor $\epsilon=0.3$, and study the learning mechanism with the different memory weights $\gamma=0.99,0.8,0.5,$
and $0.1$, respectively. The results are shown in Figure \ref{fig:Distributed-learning-mechanism}.
We see that a large enough memory weight (e.g., $\gamma\geq0.8$) is needed to guarantee that the
mechanism converges to the ESS equilibrium. When the memory
weight is large, the noise of the local estimation by each
user can be averaged out in the long run, and hence each user can
achieve an accurate estimation of the environment. When the memory
weight is small, the most recent estimations will have a great impact on the learning. This means that the learning mechanism will over-exploit the current best channels, and get stuck in a local
optimum.

\begin{figure}
\centering
\includegraphics[scale=0.53]{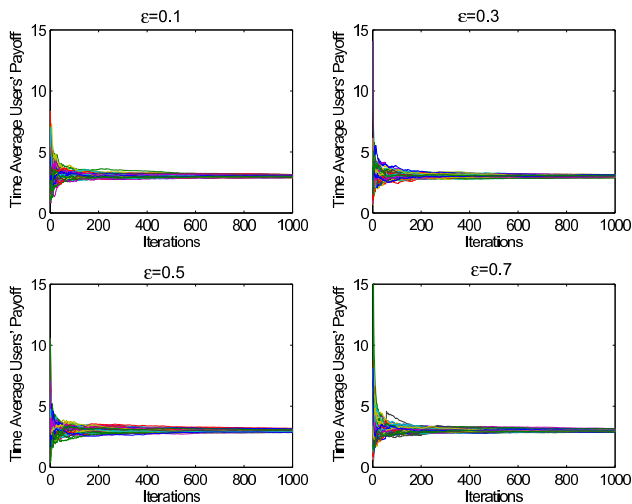}
\caption{\label{fig:Distributed-learning-mechanism-1}Distributed learning
mechanism with the memory weight $\gamma=0.99$ in the Markovian channel
environment with different dynamic factors $\epsilon$}
\end{figure}

We next set the memory weight $\gamma=0.99$, and investigate the
learning mechanism in the Markovian channel environments
with different dynamic factors $\epsilon=0.1,0.3,0.5,$ and $0.7$,
respectively. The results are shown in Figure \ref{fig:Distributed-learning-mechanism-1}.
We see that the learning mechanism can converge to the
ESS in all cases. This demonstrates that the learning
mechanism is robust to the dynamic channel state changing.

\section{\label{sec:Conclusion}Conclusion}
In this paper, we study the problem of distributed spectrum access of multiple time-varying heterogeneous licensed channels, and propose an evolutionary spectrum access mechanism based on evolutionary game theory. We show that the equilibrium of the mechanism is an evolutionarily stable strategy  and is globally stable. We further propose a learning mechanism, which requires no information exchange among the users. We show that the learning mechanism converges to the evolutionarily stable strategy on the time average. Numerical results show that the proposed mechanisms can achieve efficient and stable spectrum sharing among the users.

One possible direction of extending this result is to consider heterogeneous users, i.e. each user may achieve different mean data rates on the same channel. Another interesting direction is to take the spatial reuse effect into account. How to design an efficient evolutionarily stable spectrum access mechanism with spatial reuse will be challenging.

\section{Appendix}
\subsection{Proof of Theorem \ref{thm:For-the-evolutionary-1}} \label{proof1}
Given a population state $\boldsymbol{x}(t)=(x_{1}(t),...,x_{M}(t)),$
we divide the set of channels $\mathcal{M}$ into the following three
complete and mutually exclusive subsets: $\mathcal{M}_{1}=\{m\in\mathcal{M}|\theta_{m}B_{m}g(Nx_{m}(t))<U(\boldsymbol{x}(t))\}$,
$\mathcal{M}_{2}=\{m\in\mathcal{M}|\theta_{m}B_{m}g(Nx_{m}(t))=U(\boldsymbol{x}(t))\}$,
and $\mathcal{M}_{3}=\{m\in\mathcal{M}|\theta_{m}B_{m}g(Nx_{m}(t))>U(\boldsymbol{x}(t))\}$.

For a channel $m\in\mathcal{M}_{1}$, each user $n$ on this channel
achieves an expected payoff less than the system average payoff, i.e.,
$U_{n}(m,\boldsymbol{x}(t))=\theta_{m}B_{m}g(Nx_{m}(t))<U(\boldsymbol{x}(t))$.
According to the mechanism, each user has a probability of $\frac{\alpha}{x_{m}(t)}\left(1-\frac{U_{n}(m,\boldsymbol{x}(t))}{U(\boldsymbol{x}(t))}\right)$
to move out of the channel $m$. Since $\theta_{m}B_{m}g(Nx_{m}(t))<U(\boldsymbol{x}(t))$,
it follows that $p_{m}=0$ and hence no other users will move into
this channel. Thus, the dynamics are given as\begin{eqnarray*}
\dot{x}_{m}(t) & = & -\frac{\alpha}{x_{m}(t)}\left(1-\frac{U_{n}(m,\boldsymbol{x}(t))}{U(\boldsymbol{x}(t))}\right)x_{m}(t)\\
 & = & \alpha\left(\frac{U_{n}(m,\boldsymbol{x}(t))}{U(\boldsymbol{x}(t))}-1\right),\forall m\in\mathcal{M}_{2}.\end{eqnarray*}

For a channel $m\in\mathcal{M}_{2}$, we have $U_{n}(m,\boldsymbol{x}(t))=U(\boldsymbol{x}(t))$
and $p_{m}=0$. Thus, $\dot{x}_{m}(t)=0$, which satisfies the
conclusion.

For a channel $m\in\mathcal{M}_{3}$, each user $n$ on this channel
achieves an expected payoff higher than the system average payoff, i.e.,
$U_{n}(m,\boldsymbol{x}(t))=\theta_{m}B_{m}g(Nx_{m}(t))>U(\boldsymbol{x}(t))$.
According to the mechanism, no users will move out of the channel
$m$. Since $p_{m}>0$, there will be some other users from the channel
$m'\in\mathcal{M}_{1}$ moving into this channel. Let $\omega$ be
the fraction of population that carries out the movement. We have\begin{align*}
\omega = & \sum_{m'\in\mathcal{M}_{1}}\left(x_{m'}(t)-x_{m'}(t+1)\right)\\
= & \sum_{m'\in\mathcal{M}_{1}}\alpha\left(1-\frac{U_{n}(m',\boldsymbol{x}(t))}{U(\boldsymbol{x}(t))}\right)\\
= & \sum_{m'\in\mathcal{M}_{1}}\frac{\alpha}{U(\boldsymbol{x}(t))}\left(U(\boldsymbol{x}(t))-U_{n}(m',\boldsymbol{x}(t))\right)\\
= & \frac{\alpha}{U(\boldsymbol{x}(t))}\sum_{m'\in\mathcal{M}_{1}}\left(U(\boldsymbol{x}(t))-\theta_{m'}B_{m'}g(Nx_{m'}(t))\right).\end{align*}
Since $\theta_{m'}B_{m'}g(Nx_{m'}(t))=U(\boldsymbol{x}(t))$ for each channel $m'\in\mathcal{M}_{2}$,
and $\sum_{m'\in\mathcal{M}}\left(U(\boldsymbol{x}(t))-\theta_{m'}B_{m'}g(Nx_{m'}(t))\right)=0$,
we then obtaion\begin{eqnarray*}
 & \sum_{m'\in\mathcal{M}_{1}}\left(U(\boldsymbol{x}(t))-\theta_{m'}B_{m'}g(Nx_{m'}(t))\right)\\
= & \sum_{m'\in\mathcal{M}_{3}}\left(\theta_{m'}B_{m'}g(Nx_{m'}(t))-U(\boldsymbol{x}(t))\right).\end{eqnarray*}
Then, the fraction of the population moving into a channel $m\in\mathcal{M}_{3}$
thus is\begin{align*}
 & \dot{x}_{m}(t)  = p_{m}\omega\\
 = & \frac{\left(\theta_{m}B_{m}g(Nx_{m}(t))-U(\boldsymbol{x}(t))\right)}{\sum_{m'\in\mathcal{M}_{3}}\left(\theta_{m'}B_{m'}g(Nx_{m'}(t))-U(\boldsymbol{x}(t))\right)}\\
  & \times\frac{\alpha}{U(\boldsymbol{x}(t))}\sum_{m'\in\mathcal{M}_{3}}\left(\theta_{m'}B_{m'}g(Nx_{m'}(t))-U(\boldsymbol{x}(t))\right)\\
 = & \alpha\left(\frac{U_{n}(m,\boldsymbol{x}(t))}{U(\boldsymbol{x}(t))}-1\right),\forall m\in\mathcal{M}_{3}.\end{align*}
This completes the proof. \qed

\subsection{Proof of Theorem 2}\label{proof22}
First, it is easy to check that when $x_{m}(t)=x_{m}^{*}=\frac{\theta_{m}B_{m}}{\sum_{i=1}^{M}\theta_{i}B_{i}}$,
we have $\dot{x}_{m}(t)=0$ in the evolutionary dynamics (\ref{eq:12}). Thus,
$\boldsymbol{x}(t)$ is the equilibrium of the evolutionary dynamics.

We then apply Lyapunov's second method \cite{key-sas} to prove that the equilibrium
$\boldsymbol{x}^{*}$ is globally asymptotic stable. We use the
following Lyapunov function $V(\boldsymbol{x}(t))=-\sum_{m=1}^{M}x_{m}^{*}\ln\frac{x_{m}(t)}{x_{m}^{*}}$.
By Jensen's inequality, we first have for any $\boldsymbol{x}(t)\neq\boldsymbol{x}^{*}$
\begin{eqnarray*}
V(\boldsymbol{x}(t)) & > & -\ln\left(\sum_{m=1}^{M}x_{m}^{*}\frac{x_{m}(t)}{x_{m}^{*}}\right)\\
 & = & -\ln\left(\sum_{m=1}^{M}x_{m}(t)\right)=0.\end{eqnarray*}
Thus, we obtain that $V(\boldsymbol{x}^{*})=0$ and $V(\boldsymbol{x}(t))>0$
for any $\boldsymbol{x}(t)\neq\boldsymbol{x}^{*}$.

We then consider the time derivative of $V(\boldsymbol{x}(t))$ as\begin{align*}
   & \frac{dV(\boldsymbol{x}(t))}{dt}\\
  = & -\sum_{m=1}^{M}\frac{\partial V(\boldsymbol{x}(t))}{\partial x_{m}(t)}\dot{x}_{m}(t)=-\sum_{m=1}^{M}\frac{x_{m}^{*}}{x_{m}(t)}\dot{x}_{m}(t)\\
  = & -\frac{\alpha}{\frac{1}{M}\sum_{i=1}^{M}\frac{\theta_{i}B_{i}}{x_{i}(t)}}\sum_{m=1}^{M}\frac{x_{m}^{*}}{x_{m}(t)}\left(\frac{\theta_{m}B_{m}}{x_{m}(t)}-\frac{1}{M}\sum_{i=1}^{M}\frac{\theta_{i}B_{i}}{x_{i}(t)}\right)\\
  = & -\frac{\alpha}{M\left(\sum_{j=1}^{M}\theta_{j}B_{j}\right)\left(\frac{1}{M}\sum_{i=1}^{M}\frac{\theta_{i}B_{i}}{x_{i}(t)}\right)}\\
   & \times\left(M\sum_{m=1}^{M}\left(\frac{\theta_{m}B_{m}}{x_{m}(t)}\right)^{2}-\sum_{m=1}^{M}\sum_{i=1}^{M}\frac{\theta_{m}B_{m}}{x_{m}(t)}\frac{\theta_{i}B_{i}}{x_{i}(t)}\right)\\
  = & -\frac{\alpha}{\left(\sum_{j=1}^{M}\theta_{j}B_{j}\right)\left(\sum_{i=1}^{M}\frac{\theta_{i}B_{i}}{x_{i}(t)}\right)}\\
 &\times\sum_{m=1}^{M}\sum_{i=1}^{M}\left(\frac{\theta_{m}B_{m}}{x_{m}(t)}-\frac{\theta_{i}B_{i}}{x_{i}(t)}\right)^{2}.\end{align*}
Thus, we must have that $\frac{dV(\boldsymbol{x}^{*})}{dt}=0$ and
$\frac{dV(\boldsymbol{x}(t))}{dt}<0$ for any $\boldsymbol{x}(t)\neq\boldsymbol{x}^{*},$
which completes the proof. \qed

According to \cite{key-10}, any strict Nash equilibrium is also an ESS and hence the equilibrium $\boldsymbol{x}^{*}$ is an ESS. \qed

\subsection{Proof of Theorem 4}\label{proof4}
According to Lyapunov's second method \cite{key-sas}, we prove the global asymptotic stability by using the following Lyapunov function $V(\boldsymbol{x}(t))=L^{*}-L(\boldsymbol{x}(t))$. Since $L^{*}$ is the unique global
maximum of $L(\boldsymbol{x}(t))$ achieved at $\boldsymbol{x}^{*}$, we thus have\begin{eqnarray*}
V(\boldsymbol{x}(t)) & > & 0,\forall\boldsymbol{x}(t)\neq\boldsymbol{x}^{*},\\
V(\boldsymbol{x}^{*}) & = & 0.\end{eqnarray*}
Then differentiating $V(\boldsymbol{x}(t))$ with respective to time
$t$, we have\begin{eqnarray*}
 &  & \dot{V}(\boldsymbol{x}(t))\\
 & = & -\sum_{m=1}^{M}B_{m}g(Nx_{m})\dot{x}_{m}(t)\\
 & = & -\sum_{m=1}^{M}U_{n}(m,\boldsymbol{x}(t))\dot{x}_{m}(t)\\
 & = & -\sum_{m=1}^{M}U_{n}(m,\boldsymbol{x}(t))\frac{\alpha}{U(\boldsymbol{x}(t))}\left(U_{n}(m,\boldsymbol{x}(t))-U(\boldsymbol{x}(t))\right)\\
 & = & -\frac{\alpha}{MU(\boldsymbol{x}(t))}\left(M\sum_{m=1}^{M}\left(U_{n}(m,\boldsymbol{x}(t))\right)^{2}\right.\\
 &  & \left.-\sum_{m=1}^{M}U_{n}(m,\boldsymbol{x}(t))\sum_{m'=1}^{M}U_{n}(m',\boldsymbol{x}(t))\right)\\
 & = & -\frac{\alpha}{MU(\boldsymbol{x}(t))}\sum_{m=1}^{M}\sum_{m'=1}^{M}\left(U_{n}(m,\boldsymbol{x}(t))-U_{n}(m',\boldsymbol{x}(t))\right)^{2}\end{eqnarray*}
Thus, we obtain that\begin{eqnarray*}
\dot{V}(\boldsymbol{x}(t)) & < & 0,\forall\boldsymbol{x}(t)\neq\boldsymbol{x}^{*},\\
\dot{V}(\boldsymbol{x}^{*}) & = & 0,\end{eqnarray*}
which completes the proof. \qed

\subsection{Proof of Theorem \ref{thm:For-the-evolutionary}}\label{proof3}
We first show that the solution in (\ref{eq:aess}) is an equilibrium for the evolution
dynamics in (\ref{eq:ESS-1}). Since $U_{n}(m,\boldsymbol{x}^{*})=U_{n}(m',\boldsymbol{x}^{*})$
for any $m,m'\in\mathcal{M}$, it follows that $U(\boldsymbol{x}^{*})=\frac{1}{M}\sum_{i=1}^{M}U_{n}(i,\boldsymbol{x}^{*})=U_{n}(m,\boldsymbol{x}^{*})$
for any $m\in\mathcal{M}.$ Hence $\dot{x}_{m}=\alpha (\frac{U_{n}(m,\boldsymbol{x}^{*})}{U(\boldsymbol{x}^{*})}-1)=0$,
which is an equilibrium for the evolution dynamics in (\ref{eq:ESS-1}).

We next show that the equilibrium $\boldsymbol{x}^{*}$ is a strict
Nash equilibrium. The expected payoff
of a user $n\in\mathcal{N}$ at the equilibrium population state $\boldsymbol{x}^{*}$
is given by $U_{n}(a_{n}^{*},\boldsymbol{x}^{*}) =  \theta_{a_{n}^{*}}B_{a_{n}^{*}}g(Nx_{a_{n}^{*}}^{*}),$
where $a_{n}^{*}$ is the channel chosen by user $n$ in the population
state $\boldsymbol{x}^{*}$.
Now suppose that user $n$ makes an unilateral
deviation to another channel $a_{n}\neq a_{n}^{*}$, and the population state becomes $\boldsymbol{x}'=(x_{1}^{*},...,x_{a_{n}^{*}-1}^{*},x_{a_{n}^{*}}^{*}-\frac{1}{N},x_{a_{n}^{*}+1}^{*},...,x_{a_{n}-1}^{*},x_{a_{n}}^{*}+\frac{1}{N},x_{a_{n}+1}^{*},...,x_{N}^{*})$. Then its expected
payoff becomes $U_{n}(a_{n},\boldsymbol{x}')=\theta_{a_{n}}B_{a_{n}}g(Nx_{a_{n}}^{*}+1)<\theta_{a_{n}}B_{a_{n}}g(Nx_{a_{n}}^{*}).$
For the equilibrium $\boldsymbol{x}^{*}$, we have $ \theta_{a_{n}^{*}}B_{a_{n}^{*}}g(Nx_{a_{n}^{*}}^{*})  =  \theta_{a_{n}}B_{a_{n}}g(Nx_{a_{n}}^{*}).$
It follows that $
U_{n}(a_{n}^{*},\boldsymbol{x}^{*})>U_{n}(a_{n},\boldsymbol{x}'),\forall a_{n}\neq a_{n}^{*},n\in\mathcal{N}$,
which is a strict Nash equilibrium.

\subsection{Proof of Theorem \ref{lm2}}\label{proof2}
The key idea of the proof is to first obtain the discrete time dynamics of the learning mechanism, and then derive the corresponding mean continuous time dynamics.

For simplicity, we first define that \[
A_{m,n}(T)=\sum_{\tau=0}^{T-1}\gamma^{T-\tau-1}Z_{m,n}(\tau)\]
and\[
C_{n}(T)=\frac{\sum_{t=1}^{t_{\max}}b_{i}(t)I_{\{a_{n}(t,T)=a_{n}(T)\}}}{t_{\max}},\]
 where $I_{\{a_{n}(t,T)=a_{n}(T)\}}$ indicates whether user $n$
successfully grabs the chosen channel $a_{n}(T)$, and hence $C_{n}(T)$
denotes the average throughput it received at period $T$. According
to (\ref{est2}) and (\ref{est1}), we have\begin{align*}
  & A_{m,n}(T+1) = Z_{m,n}(T)+\gamma\sum_{\tau=0}^{T-1}\gamma^{T-\tau-1}Z_{m,n}(\tau)\\
  = & (1-\gamma)C_{n}(T)I_{\{a_{n}(T)=m\}}+(1-\gamma)\sum_{\tau=0}^{T-1}\gamma^{T-\tau-1}Z_{m,n}(\tau)\\
   & +\gamma\sum_{\tau=0}^{T-1}\gamma^{T-\tau-1}Z_{m,n}(\tau)\\
 = & (1-\gamma)C_{n}(T)I_{\{a_{n}(T)=m\}}+\sum_{\tau=0}^{T-1}\gamma^{T-\tau-1}Z_{m,n}(\tau)\\
  = & (1-\gamma)C_{n}(T)I_{\{a_{n}(T)=m\}}+A_{m,n}(T),\end{align*}
where $I_{\{a_{n}(T)=m\}}$ indicates whether user $n$ chooses channel
$m$ in period $T$. Then the mixed strategy update in (\ref{eq:ESL-2}) becomes
\begin{align*}
& f_{m,n}(T+1) = \frac{A_{m,n}(T+1)}{\sum_{i=1}^{M}A_{i,n}(T+1)}\\
 = & \frac{A_{m,n}(T)+(1-\gamma)C_{n}(T)I_{\{a_{n}(T)=m\}}}{\sum_{i=1}^{M}A_{i,n}(T)+(1-\gamma)C_{n}(T)},\end{align*}
For the chosen channel $j$ (i.e., $I_{\{a_{n}(T)=j\}}=1$), we further
have\begin{align}
 & f_{j,n}(T+1) = \frac{A_{j,n}(T)}{\sum_{i=1}^{M}A_{i,n}(T)+(1-\gamma)C_{n}(T)}\\ \nonumber
  & +\frac{(1-\gamma)C_{n}(T)}{\sum_{i=1}^{M}A_{i,n}(T)+(1-\gamma)C_{n}(T)}\\ \nonumber
  = & \frac{A_{j,n}(T)}{\sum_{i=1}^{M}A_{i,n}(T)}\frac{\sum_{i=1}^{M}A_{i,n}(T)}{\sum_{i=1}^{M}A_{i,n}(T)+(1-\gamma)C_{n}(T)}\\ \nonumber
   & +\frac{(1-\gamma)C_{n}(T)}{\sum_{i=1}^{M}A_{i,n}(T)+(1-\gamma)C_{n}(T)}\\ \nonumber
  = & f_{j,n}(T)\left(1-\frac{(1-\gamma)C_{n}(T)}{\sum_{i=1}^{M}A_{i,n}(T)+(1-\gamma)C_{n}(T)}\right)\\  
  & +\frac{(1-\gamma)C_{n}(T)}{\sum_{i=1}^{M}A_{i,n}(T)+(1-\gamma)C_{n}(T)}. \label{eq:33}\end{align}
Let $\beta(T)=\frac{(1-\gamma)}{\sum_{i=1}^{M}A_{i,n}(T)+(1-\gamma)C_{n}(T)}$, and (\ref{eq:33}) can be expresses as
\begin{equation}
f_{j,n}(T+1)=f_{j,n}(T)(1-\beta(T)C_{n}(T))+\beta(T)C_{n}(T).\label{eq:p1}\end{equation}
Similarly, for the unchosen channel $j'$(i.e., $I_{\{a_{n}(T)=j'\}}=0$),
we have\begin{equation}
f_{j',n}(T+1)=f_{j',n}(T)(1-\beta(T)C_{n}(T)).\label{eq:p2}\end{equation}
According to (\ref{eq:p1}) and (\ref{eq:p2}), we thus obtain the discrete time
learning dynamics as\begin{align}
 f_{m,n}(T+1)-f_{m,n}(T)=\beta(T)C_{n}(T)(I_{\{a_{n}(T)=m\}}-f_{m,n}(T)).\label{eq:p3}\end{align}

Since $\beta(T)\rightarrow0$ as $\gamma\rightarrow1$,  by the theory of stochastic approximation (Theorem 3.2) in \cite{kushner1984approximation}, the limiting behavior of the stochastic difference
equations in (\ref{eq:p3}) is the same as its mean continuous time
dynamics by taking the expectation on RHS of (\ref{eq:p3}) with respective
to $\boldsymbol{f}(T)$, i.e., \begin{align}
  & \dot{f}_{m,n}(T) = E[C_{n}(T)(I_{\{a_{n}(T)=m\}}-f_{m,n}(T))|\boldsymbol{f}(T)]\nonumber \\
= & (1-f_{m,n}(T))E[C_{n}(T)|a_{n}(t)=m,\boldsymbol{f}(T)]f_{m,n}(T)\nonumber \\
 & -f_{m,n}(T)\sum_{i\ne m}E[C_{n}(T)|a_{n}(t)=i,\boldsymbol{f}(T)]f_{i,n}(T).\label{eq:p4}\end{align}
Since $C_{n}(T)$ is the sample averaging estimation of the expected throughput of the chosen channel, by the central
limit theorem, we have $E[C_{n}(T)|a_{n}(t)=m,\boldsymbol{f}(T)]=E[U_{n}(a_{n}(t)=m,\boldsymbol{x}(T))|\boldsymbol{f}(T)]=Q_{m,n}(\boldsymbol{f}(T))$.
Then the mean dynamics in (\ref{eq:p4}) can be written as\begin{align*}
 \dot{f}_{m,n}(T) & = Q_{m,n}(\boldsymbol{f}(T))(1-f_{m,n}(T))f_{m,n}(T)\\
  & -f_{m,n}(T)\sum_{i\ne m}Q_{i,n}(\boldsymbol{f}(T))f_{i,n}(T)\\
 = & f_{m,n}(T)\left(Q_{m,n}(\boldsymbol{f}(T))-\sum_{i=1}^{M}Q_{i,n}(\boldsymbol{f}(T))f_{i,n}(T)\right),\end{align*}
which completes the proof. \qed

\subsection{Proof of Theorem \ref{thm:haha}}\label{proof6}
We first denote the following function \[
\Phi(\boldsymbol{f}(T))=E[\sum_{i=1}^{M}\int_{-\infty}^{x_{i}(T)}\theta_{i}B_{i}g(Nz)dz|\boldsymbol{f}(T)],\]
and \[
\Phi_{m,n}(\boldsymbol{f}(T))=E[\sum_{i=1}^{M}\int_{-\infty}^{x_{i}(T)}\theta_{i}B_{i}g(Nz)dz|a_{n}(T)=m,\boldsymbol{f}(T)].\]
Obviously, we have \begin{equation*}
\Phi(\boldsymbol{f}(T))=\sum_{m=1}^{M}\Phi_{m,n}(\boldsymbol{f}(T))f_{m,n}(T).\end{equation*}
We further denote $\boldsymbol{x}_{-n}(T)\triangleq\left(x_{m}^{-n}(T),m\in\mathcal{M}\right)$
as the population state of all other users without user $n$. By considering the user distributions on the chosen channel $m$ by user $n$ and the other channels, we then
have\begin{align}
   & \Phi_{m,n}(\boldsymbol{f}(T))\nonumber \\
  = & E[\sum_{i=1}^{M}\int_{-\infty}^{x_{i}(T)}\theta_{i}B_{i}g(Nz)dz|a_{n}(T)=m,\boldsymbol{f}(T)]\nonumber \\
  = & \sum_{\boldsymbol{x}_{-n}(T)}E[\sum_{i\neq m}\int_{-\infty}^{x_{i}^{-n}(T)}\theta_{i}B_{i}g(Nz)dz\nonumber \\
   & +\int_{-\infty}^{x_{m}^{-n}(T)+\frac{1}{N}}\theta_{m}B_{m}g(Nz)dz|a_{n}(T)=m,\boldsymbol{x}_{-n}(T)]\nonumber \\
   & \times Pr\{\boldsymbol{x}_{-n}(T)|\boldsymbol{f}(T)\}\nonumber \\
  = & \sum_{\boldsymbol{x}_{-n}(T)}\left(\sum_{i\neq m}\int_{-\infty}^{x_{i}^{-n}(T)}\theta_{i}B_{i}g(Nz)dz\right.\nonumber \\
   & \left.+\int_{-\infty}^{x_{m}^{-n}(T)+\frac{1}{N}}\theta_{m}B_{m}g(Nz)dz\right)Pr\{\boldsymbol{x}_{-n}(T)|\boldsymbol{f}(T)\}.\label{eq:1-1}\end{align}
Similarly, we can obtain\begin{align}
  & \Phi_{m',n}(\boldsymbol{f}(T)) = \sum_{\boldsymbol{x}_{-n}(T)}\left(\sum_{i\neq m'}\int_{-\infty}^{x_{i}^{-n}(T)}\theta_{i}B_{i}g(Nz)dz\right.\nonumber \\
   & \left.+\int_{-\infty}^{x_{m'}^{-n}(T)+\frac{1}{N}}\theta_{m'}B_{m'}g(Nz)dz\right)Pr\{\boldsymbol{x}_{-n}(T)|\boldsymbol{f}(T)\}.\label{eq:1-2}\end{align}
It follows that \begin{align}
   & \Phi_{m,n}(\boldsymbol{f}(T))-\Phi_{m',n}(\boldsymbol{f}(T))\nonumber \\
  = & \sum_{\boldsymbol{x}_{-n}(T)}\left(\sum_{i\neq m}\int_{-\infty}^{x_{i}^{-n}(T)}\theta_{i}B_{i}g(Nz)dz\right.\nonumber \\
   & \left.+\int_{-\infty}^{x_{m}^{-n}(T)+\frac{1}{N}}\theta_{m}B_{m}g(Nz)dz\right)Pr\{\boldsymbol{x}_{-n}(T)|\boldsymbol{f}(T)\}\nonumber \\
   & -\sum_{\boldsymbol{x}_{-n}(T)}\left(\sum_{i\neq m'}\int_{-\infty}^{x_{i}^{-n}(T)}\theta_{i}B_{i}g(Nz)dz\right.\nonumber \\
   & \left.+\int_{-\infty}^{x_{m'}^{-n}(T)+\frac{1}{N}}\theta_{m'}B_{m'}g(Nz)dz\right)Pr\{\boldsymbol{x}_{-n}(T)|\boldsymbol{f}(T)\}\nonumber \\
  = & \sum_{\boldsymbol{x}_{-n}(T)}\left(\int_{-\infty}^{x_{m}^{-n}(T)+\frac{1}{N}}\theta_{m}B_{m}g(Nz)dz\right.\nonumber \\
  & \left.-\int_{-\infty}^{x_{m}^{-n}(T)}\theta_{m}B_{m}g(Nz)\right)Pr\{\boldsymbol{x}_{-n}(T)|\boldsymbol{f}(T)\}\nonumber \\
  & -\sum_{\boldsymbol{x}_{-n}(T)}\left(\int_{-\infty}^{x_{m'}^{-n}(T)+\frac{1}{N}}\theta_{m'}B_{m'}g(Nz)dz\right.\nonumber \\
  & \left.-\int_{-\infty}^{x_{m'}^{-n}(T)}\theta_{m'}B_{m'}g(Nz)\right)Pr\{\boldsymbol{x}_{-n}(T)|\boldsymbol{f}(T)\}.\label{eq:lss}\end{align}
Since $N$ is large, we obtain that for $i\in\{m,m'\}$ \begin{align}
   & \int_{-\infty}^{x_{i}^{-n}(T)+\frac{1}{N}}\theta_{i}B_{i}g(Nz)dz-\int_{-\infty}^{x_{i}^{-n}(T)}\theta_{i}B_{i}g(Nz)\nonumber \\
  = & \int_{x_{i}^{-n}(T)}^{x_{i}^{-n}(T)+\frac{1}{N}}\theta_{i}B_{i}g(Nz)dz = \int_{Nx_{i}^{-n}(T)}^{Nx_{i}^{-n}(T)+1}\theta_{i}B_{i}g(z)dz\nonumber \\
  \approx & \theta_{i}B_{i}g(Nx_{i}^{-n}(T)+1).\label{eq:lss-2-1}\end{align}
According to (\ref{eq:lss}) and (\ref{eq:lss-2-1}), we have\begin{align}
    & \Phi_{m,n}(\boldsymbol{f}(T))-\Phi_{m',n}(\boldsymbol{f}(T))\nonumber \\
  = & \sum_{\boldsymbol{x}_{-n}(T)}\left(\theta_{m}B_{m}g(Nx_{m}^{-n}(T)+1)\right.\nonumber \\
   & -\left.\theta_{m'}B_{m'}g(Nx_{m'}^{-n}(T)+1)\right)Pr\{\boldsymbol{x}_{-n}(T)|\boldsymbol{f}(T)\}\nonumber \\
   = & E[U_{n}(a_{n}=m,\boldsymbol{x}(t))|\boldsymbol{f}(T)]-E[U_{n}(a_{n}=m',\boldsymbol{x}(t))|\boldsymbol{f}(T)]\nonumber \\
 = & Q_{m,n}(\boldsymbol{f}(T))-Q_{m',n}(\boldsymbol{f}(T)).\label{eq:lss-2}\end{align}

We then consider the variation of $\Phi(\boldsymbol{f}(T))$ along
the trajectories of learning dynamics in (\ref{learnD}), i.e., differentiating
$\Phi(\boldsymbol{f}(T))$ with respective to time $T$, \begin{align}
 & \frac{d\Phi(\boldsymbol{f}(T))}{dT} = \sum_{m=1}^{M}\frac{d\Phi(\boldsymbol{f}(T))}{df_{m,n}(T)}\frac{df_{m,n}(T)}{dT} \nonumber \\
 = & \sum_{m=1}^{M}\Phi_{m,n}(\boldsymbol{f}(T))f_{m,n}(T)\nonumber \\
  & \times\left(Q_{m,n}(\boldsymbol{f}(T))-\sum_{i=1}^{M}f_{i,n}(T)Q_{i,n}(\boldsymbol{f}(T))\right)\nonumber \\
  = & \frac{1}{2}\sum_{m=1}^{M}\sum_{i=1}^{M}f_{m,n}(T)f_{i,n}(T)\nonumber \\
   & \times\left(Q_{m,n}(\boldsymbol{f}(T))-Q_{i,n}(\boldsymbol{f}(T))\right)\left(\Phi_{m,n}(\boldsymbol{f}(T))-\Phi_{i,n}(\boldsymbol{f}(T))\right)\nonumber \\
  = & \frac{1}{2}\sum_{i=1}^{M}\sum_{j=1}^{M}f_{i,n}(T)f_{j,n}(T)\left(Q_{i,n}(\boldsymbol{f}(T))-Q_{j,n}(\boldsymbol{f}(T))\right)^{2}\nonumber \\
 \ge & 0.\label{eq:s4}\end{align}
Hence $\Phi(\boldsymbol{f}(T))$ is non-decreasing along the trajectories
of the ODE (\ref{learnD}). According to Theorem 2.7 in \cite{key-sas}, the learning mechanism asymptotically
converges to a limit point $\boldsymbol{f}^{*}$ such that \begin{equation}
\frac{d\Phi(\boldsymbol{f}^{*})}{dT}=0,\label{eq:pr5}\end{equation}
i.e., for any $m,i\in\mathcal{M},n\in\mathcal{N}$\begin{equation}
f_{m,n}^{*}f_{i,n}^{*}\left(Q_{m,n}(\boldsymbol{f}^{*})-Q_{i,n}(\boldsymbol{f}^{*})\right)=0.\label{eq:pr8}\end{equation}
According to the mixed strategy update in (\ref{eq:ESL-2}), we know that $f_{m,n}(T)>0$ for any $m\in\mathcal{M}$.
Thus, from (\ref{eq:pr8}), we must have $Q_{m,n}(\boldsymbol{f}^{*})=Q_{i,n}(\boldsymbol{f}^{*})$. \qed

\bibliographystyle{ieeetran}
\bibliography{Evolution}

\end{document}